\documentclass[12pt]{article}

\usepackage{authblk}
\usepackage{amsmath}
\usepackage{graphicx}
\usepackage{enumerate}
\usepackage{natbib}
\usepackage{url}
\usepackage{tcolorbox}

\usepackage{enumitem}
\usepackage{amsmath}    % for general math formatting
\usepackage{amssymb}    % for \mathbb and other math symbols
\usepackage{amsthm}     % for theorem environments

\theoremstyle{plain}

\theoremstyle{definition}

\usepackage{caption}
\usepackage{graphicx}
\usepackage{xcolor}
\usepackage{algorithm}
\usepackage{algorithmic}
\usepackage{datetime2} % for \DTMnow

\usepackage{tikz}
\usetikzlibrary{positioning,arrows.meta}
\usepackage{booktabs}  % Required for \toprule, \midrule, \bottomrule

\begin{document}

\setlength{\parskip}{0.4\baselineskip plus 2pt}
\setlength{\parindent}{0pt}

\def\spacingset#1{\renewcommand{\baselinestretch}%
{#1}\small\normalsize} \spacingset{1}

\title{\bf Adaptive Geometric Regression for High-Dimensional Structured Data\thanks{Research reported in this publication was supported by grant number INV-048956 from the Gates Foundation.}}
\author{Pawel Gajer$^{1}$\thanks{Corresponding author: pgajer@gmail.com}\; and Jacques Ravel$^{1}$}
\date{%
\small
$^1$University of Maryland School of Medicine
}
\maketitle

{\renewcommand{\thefootnote}{}\footnotetext{Software: The methods are implemented in the R package \texttt{gflow}, available at \url{https://github.com/pgajer/gflow}.}}

\begin{abstract}
  We present a geometric framework for regression on structured high-dimensional
  data that shifts the analysis from the ambient space to a geometric object
  capturing the data's intrinsic structure. The method addresses a fundamental
  challenge in analyzing datasets with high ambient dimension but low intrinsic
  dimension, such as microbiome compositions, where traditional approaches fail
  to capture the underlying geometric structure. Starting from a k-nearest
  neighbor covering of the feature space, the geometry evolves iteratively
  through heat diffusion and response-coherence modulation, concentrating mass
  within regions where the response varies smoothly while creating diffusion
  barriers where the response changes rapidly. This iterative refinement
  produces conditional expectation estimates that respect both the intrinsic
  geometry of the feature space and the structure of the response.
\end{abstract}


\section{Introduction}

Modern datasets present a statistical paradox: they possess hundreds or
thousands of features (high ambient dimension) yet their intrinsic dimension is
often orders of magnitude lower \cite{fefferman2016testing, levina2004maximum}.
Classical approaches such as LASSO regression perform analysis in the ambient
space, imposing parametric constraints including sparsity and linearity
assumptions \cite{tibshirani1996regression, hastie2015statistical}. Machine
learning methods achieve greater flexibility by adapting to data structure, yet
this complicates inferential analysis. Recent work demonstrates that this
flexibility stems from implicit adaptation to the complex nonlinear geometry
underlying the data \cite{gajer2025geometry}, raising a natural question: Can
such low-dimensional intrinsic geometry be exploited explicitly for principled
inferential regression analysis?

We present a geometric framework that addresses this question by shifting
regression analysis from the ambient space to learned geometric objects
capturing the data's intrinsic structure. While our motivating application comes
from microbiome studies \cite{fettweis2019vaginal, callahan2016dada2}, the
framework applies broadly to any domain where data possess low-dimensional
structure embedded in high-dimensional space. Rather than fixing the regression
domain in advance, our approach iteratively constructs a triple
$(K^{(j)}, g^{(j)}, \rho^{(j)})$, where $K^{(j)}$ is a simplicial complex
encoding neighborhood relationships, $g^{(j)}$ is a Riemannian structure
assigning inner products to chains, and
$\rho^{(j)} = \{\rho^{(j)}_p\}_{p \geq 0}$ represents density estimates on
$p$-simplices of different dimensions. The geometry and density evolve jointly
through heat diffusion and response-coherence modulation, producing conditional
expectation estimates that respect both the intrinsic geometry of features and
the structure of the response.

Classical kernel density estimation and kernel regression provide powerful
nonparametric approaches in Euclidean spaces, smoothing empirical measures by
convolving with kernels such as the Gaussian \cite{silverman1986density,
  scott2015multivariate, wand1994kernel}. These methods achieve minimax optimal
rates in low dimensions but become impractical beyond dimension 10 to 20 due to
the curse of dimensionality \cite{wasserman2006all, li2007nonparametric}: the
convergence rate $O(n^{-4/(4+d)})$ with optimal bandwidth implies that sample
size requirements grow exponentially with dimension $d$ \cite{stone1982optimal,
  tsybakov2009introduction}. Graph-based analogues of kernel methods offer one
escape route by defining Laplacian operators on proximity graphs, typically
k-nearest neighbor (k-NN) graphs, and using heat kernels as smoothing mechanisms
\cite{belkin2006manifold, coifman2006diffusion, chung1997spectral}. This
approach reduces effective dimensionality from the ambient space dimension to
the intrinsic dimension of the data manifold, and empirically performs well when
the underlying structure is lower-dimensional \cite{hein2007graph,
  vonluxburg2008consistency}.

Both classical kernel methods and graph-based extensions share a fundamental
limitation: they assume the analysis domain is fixed in advance, either as the
Euclidean space in which observations are embedded or as a graph constructed
once from pairwise similarities. In many applications, fixing the domain at the
outset leads to instability. Consider predicting spontaneous preterm birth from
vaginal microbiome data, where bacterial abundance profiles determine pregnancy
outcomes \cite{fettweis2019vaginal, elovitz2019cervicovaginal,
  callahan2017replication}. A k-NN graph built solely from Euclidean or
compositional distances cannot distinguish regions where pregnancy outcomes vary
smoothly from regions where outcomes change abruptly. Two samples may be
Euclidean neighbors yet belong to different risk strata if microbial composition
differs in outcome-relevant ways. Conversely, samples separated in feature space
may share outcome coherence if the response varies smoothly along connecting
paths. This disconnect between topology (which points are neighbors) and the
prediction target (which regions have similar outcomes) hampers downstream
inference, particularly when identifying outcome-associated subpopulations or
performing stable segmentation of feature space \cite{rinaldo2010stability,
  chaudhuri2014rates, gerber2012morse}.

The construction proceeds inductively. Given current densities $\rho^{(j)}$, we
update the Riemannian structure $g^{(j)}$, assemble the associated Hodge
Laplacians $L^{(j)} = \{L^{(j)}_p\}$, and compute new densities via heat
diffusion,
\begin{equation}
\rho^{(j+1)}_p = e^{-t L^{(j)}_p} \rho^{(j)}_p.
\end{equation}
When a response variable is present, we smooth it via spectral filtering
$\hat{y}^{(j)} = \exp(-\eta L^{(j)}_0)y$ with regularization parameter $\eta$
selected by generalized cross-validation, then modulate edge masses using a
penalty function that down-weights edges spanning large response differences
\cite{zhou2004learning, zhu2003semi}. This coupling between density, geometry,
and response yields a self-consistent refinement where the domain and
distributions on it evolve together until reaching equilibrium.

To understand the construction, we begin with familiar tools. Given a dataset
$X = \{x_1, \ldots, x_n\}$, a k-NN graph $G_k(X)$ records pairwise neighbor
relationships. Classical Laplacian smoothing on $G_k(X)$ yields vertex-level
estimates by diffusing mass across edges \cite{zhou2004learning,
  belkin2006manifold}. However, graphs record only pairwise relations. The
natural generalization to a simplicial complex $K(X)$ extends this by recording
not only edges but also higher-order cliques: 2-simplices (triangles) capture
triples of points sharing neighborhoods, 3-simplices (tetrahedra) capture
quadruples, and so forth \cite{edelsbrunner2010computational,
  carlsson2009topology, ghrist2008barcodes}. From a k-NN covering
$\mathcal{U} = \{\widehat{N}_k(x_i)\}$, where $\widehat{N}_k(x_i)$ denotes the
closed k-nearest-neighbor ball of $x_i$, we construct the nerve complex:
vertices correspond to samples, edges connect samples with overlapping
neighborhoods, and higher-dimensional simplices record multi-way overlaps
\cite{borsuk1948imbedding, dowker1952homology}.

Once the combinatorial structure $K(X)$ is established, we must specify how to
measure lengths, areas, and volumes. In discrete settings, a Riemannian
structure amounts to choosing inner products on chain spaces. Formally, we
assign to each chain space $C_p$ a symmetric positive definite matrix $M_p$
encoding these inner products. The collection
$g = \{M_p : 0 \leq p \leq p_{\max}\}$ constitutes the Riemannian structure on
the complex. It induces norms on chains and, via the standard Hodge formula
\cite{desbrun2005discrete, hirani2003discrete, grady2010discrete},
\begin{equation}
L_p = B_{p+1} M_{p+1}^{-1} B_{p+1}^T M_p + M_p^{-1} B_p^T M_{p-1} B_p,
\end{equation}
determines the Hodge Laplacian $L_p$ on $p$-chains, where $B_p$ denotes the
boundary operator. For vertex chains ($p=0$), this reduces to the weighted graph
Laplacian commonly used in spectral clustering and manifold learning
\cite{vonluxburg2007tutorial, belkin2003laplacian}.

The key innovation is that $g^{(j)}$ adapts to both data geometry and response
structure. Given a Riemannian structure $g^{(j)}$ encoded by vertex masses
$\rho^{(j)}_0$ and edge masses $\rho^{(j)}_1$, the graph Laplacian $L^{(j)}_0$
generates diffusion. The heat kernel $K_t^{(j)} = \exp(-tL^{(j)}_0)$ at time $t$
describes random walk distributions \cite{lovasz1993random, grigoryan2009heat},
and applying it to point masses yields density estimates that respect the
geometry: in regions where edges are short (small edge masses), diffusion
spreads rapidly and densities smooth aggressively, while long edges act as
barriers that slow diffusion and preserve local structure
\cite{jones2008manifold, singer2006graph}. We use evolved densities
$\rho^{(j+1)}_0$ to update the geometry by setting vertex masses proportional to
density and rescaling edge masses by density factors, which shortens edges in
dense regions \cite{lafon2006diffusion, nadler2006diffusion}. When a response
variable $y$ is present, we further modulate edge masses by outcome coherence:
edges crossing large outcome jumps receive reduced mass, creating diffusion
barriers at response boundaries while preserving corridors within
response-coherent regions.

The framework operates at two levels of geometric adaptation. At the first
level, described above, edge and vertex masses adjust within a fixed complex
structure $K^{(0)}$, with mass redistribution creating diffusion corridors in
response-coherent regions and barriers at response boundaries. At the second
level, the complex structure itself evolves using learned geometry. The current
implementation maintains the fixed complex throughout iterations, providing
efficient response-aware refinement. We now describe the complete framework that
includes complex evolution, which represents the natural theoretical completion
and a direction for future implementation.

In the complete framework, refined densities $\rho^{(j)}$ define density-aware
distance metrics between points. For example, diffusion distances computed from
the heat kernel $K_t^{(j)}$ \cite{coifman2006diffusion, richards2013geometry}
measure similarity based on heat flow patterns, placing nearby two vertices from
which heat spreads similarly across the complex. Alternatively, path-based
metrics can integrate density reciprocals along geodesics, yielding distances
that account for both geometric proximity and density variations. These evolved
distance metrics enable reconstruction of the k-NN covering
$\mathcal{N}^{(j+1)}$ and hence the nerve complex $K^{(j+1)}$ itself, yielding a
fully adaptive iteration:
\begin{equation}
\rho^{(j)} \to g^{(j)} \to L^{(j)} \to \{\hat{y}^{(j)}, \rho^{(j+1)}\} \to d^{(j+1)} \to K^{(j+1)} \to \cdots,
\end{equation}
where each arrow represents a well-defined update rule. This closes the loop
between neighborhood structure, geometry, density, and response: the complex
structure adapts to learned distances, which in turn depend on the evolved
density and response-modulated geometry. While this complete framework offers
maximal adaptivity, the fixed-complex variant ($K^{(j)} = K^{(0)}$ for all $j$)
already captures the essential geometric adaptation through mass redistribution
and provides the foundation for the current implementation.

From a statistical perspective, our methodology generalizes diffusion-based
kernel estimators. When restricted to vertices ($p = 0$) with fixed topology,
the procedure recovers graph kernel density estimation \cite{hein2007graph,
ting2011analysis}. The extension to adaptive geometry stabilizes diffusion
corridors in dense regions while suppressing spurious fluctuations in sparse
ones. Connections to machine learning partition models provide additional
intuition: methods such as decision trees, MARS, and neural networks learn
feature space partitions depending on both features $X$ and response $y$,
concentrating resolution where prediction error is high \cite{breiman1984classification,
friedman1991multivariate, goodfellow2016deep}. Each learned partition induces
a simplicial complex where edges connecting samples in dense, homogeneous cells
are short, while edges crossing decision boundaries are long. Our iterative
framework formalizes and extends this outcome-aware geometry to regression and
density estimation settings.

The framework admits natural extensions beyond the basic vertex-edge iteration.
Just as we smooth the response $y$ to obtain $\hat{y}^{(j)}$, we can smooth the
feature matrix $X$ to obtain denoised features $\hat{X}^{(j)}$, iterating
between smoothing features, recomputing geometry, and smoothing responses to
yield joint denoising and prediction models \cite{singer2009non,
  trillos2016variational, szlam2008diffusion}. Higher-dimensional refinement
extends the iteration from vertices and edges to triangles and tetrahedra, using
the full wedge-Gram construction to define Hodge Laplacians $L^{(j)}_p$ on
$p$-cochains \cite{lim2020hodge, schaub2021random}. These higher Laplacians
capture multi-way interactions and support more nuanced density estimates
accounting for the filling structure of the complex, not just its graph skeleton
\cite{barbarossa2020topological, giusti2016two}. Another direction incorporates
temporal or hierarchical structure, allowing the complex to evolve across time
slices or to organize features at multiple resolutions
\cite{petri2014homological, sizemore2018knowledge}.

The remainder of the paper is organized as follows. Section 2 establishes the
mathematical framework connecting simplicial complexes, nerve constructions, and
Riemannian structures. Section 3 presents the iterative algorithm with
initialization, density evolution, and response-coherence modulation. Section 4
addresses computational implementation and optimization. Section 5 discusses
parameter selection distinguishing automatic, geometrically motivated, and
tunable parameters. Section 6 develops extensions to multivariate responses,
classification, and uncertainty quantification.

\section*{2. Data-Driven Geometric Structures}

This section develops the geometric structures underlying our regression
framework. We begin with the familiar case of graph Laplacian smoothing
(Section 2.1), then extend to simplicial complexes that capture higher-order
relationships (Section 2.2). The central contribution is a construction that
equips these complexes with Riemannian structures derived from neighborhood
overlaps (Sections 2.3-2.4), yielding geometry that adapts naturally to data
density and response structure.

\subsection*{2.1 The Graph Case: Laplacian Smoothing as a Starting Point}

Consider a $k$-nearest neighbor graph $G_k(X)$ built from a dataset $X = \{x_1, \ldots, x_n\} \subset \mathbb{R}^d$. Assign vertex masses $m_i > 0$ to each vertex $x_i$ and edge weights $w_{ij} \geq 0$ to each edge connecting $x_i$ and $x_j$. These weights might represent local density (higher $m_i$ in dense regions) or inverse distances (smaller $w_{ij}$ for distant vertices). Collect these into diagonal matrices $M_0 = \text{diag}(m_1, \ldots, m_n)$ and a weighted adjacency matrix $W$ with entries $W_{ij} = w_{ij}$.

The graph Laplacian $L_0$ governs diffusion on the graph. In the symmetric normalized form,
$$
L_0 = M_0^{-1/2}(D - W)M_0^{-1/2},
$$
where $D = \text{diag}(\sum_j w_{ij})$ is the degree matrix. Alternatively, in the random-walk form,
$$
L_0 = M_0^{-1}(D - W).
$$
Both forms share the same eigenvectors and produce similar smoothing behavior. The heat kernel $K_t = \exp(-tL_0)$ acts as a smoothing operator: given an initial distribution $\rho^{(0)}$ (for instance, point masses at data locations), the smoothed density is
$$
\rho^{(1)} = K_t \rho^{(0)} = e^{-tL_0} \rho^{(0)}.
$$
This is the graph analogue of Gaussian kernel smoothing. The diffusion time $t$ controls the scale of smoothing, with larger $t$ producing more diffusion.

For regression, if $y \in \mathbb{R}^n$ is a response vector observed at the vertices, Tikhonov regularization yields the smoothed estimate
$$
\hat{y} = (I + \eta L_0)^{-1} y,
$$
where $\eta > 0$ controls the tradeoff between fidelity to observations and smoothness. The Laplacian penalizes rapid variation across edges weighted by $w_{ij}$: if two neighboring vertices $x_i$ and $x_j$ have large $w_{ij}$ (short edge), the regularization encourages $\hat{y}(i) \approx \hat{y}(j)$.

This graph-based framework is well-established in statistics and machine
learning. Our contribution is to extend it in two ways: first, by allowing the
geometry (the matrices $M_0$ and $W$) to evolve based on the density and
response estimates; second, by incorporating higher-dimensional simplices beyond
vertices and edges, which capture multi-way neighborhood overlaps.

\subsection*{2.2 From Coverings to Simplicial Complexes}

A graph records which pairs of points are neighbors. A simplicial complex
extends this idea by recording not only pairwise connections but also
higher-order cliques: triangles, tetrahedra, and so forth. Statistically, this
is simply a coordinate-free way to encode multi-way neighborhood relationships.

Consider a finite point cloud $X = \{x_1, \ldots, x_n\} \subset \mathbb{R}^d$ equipped with a metric $d(\cdot, \cdot)$, typically the Euclidean distance. A covering $\mathcal{U} = \{U_\alpha\}_{\alpha \in I}$ of $X$ is a collection of subsets such that every point in $X$ belongs to at least one $U_\alpha$. The most common choice in statistical applications is the $k$-nearest neighbor (kNN) covering, where we set
$$
U_i = \widehat{N}_k(x_i) = \{x \in X : x \text{ is among the } k \text{ nearest neighbors of } x_i\} \cup \{x_i\}.
$$
This covering assigns to each sample point $x_i$ the closed ball consisting of $x_i$ itself and its $k$ nearest neighbors. Two points belong to overlapping neighborhoods precisely when they are near each other in the ambient space, making the covering a local encoding of proximity structure.

From any covering $\mathcal{U}$ we construct the nerve complex $K(\mathcal{U})$, a simplicial complex that records the pattern of overlaps among the covering sets. The construction proceeds inductively by dimension. A vertex of $K(\mathcal{U})$ corresponds to each covering set $U_\alpha$. An edge $[U_\alpha, U_\beta]$ is present whenever the two sets overlap, that is, $U_\alpha \cap U_\beta \neq \varnothing$. More generally, a $p$-simplex $[U_{\alpha_0}, U_{\alpha_1}, \ldots, U_{\alpha_p}]$ is present whenever the $(p+1)$ covering sets have nonempty intersection,
$$
U_{\alpha_0} \cap U_{\alpha_1} \cap \cdots \cap U_{\alpha_p} \neq \varnothing.
$$
A 2-simplex (triangle) thus records a triple of covering sets with mutual overlap, a 3-simplex (tetrahedron) records a quadruple, and so on.

For the kNN covering $\mathcal{U} = \{\widehat{N}_k(x_i)\}_{i=1}^n$, the nerve complex is denoted $K_\bullet^k(X)$. Its vertices correspond to the $n$ sample points. An edge connects $x_i$ and $x_j$ if and only if their kNN neighborhoods overlap, that is, if $\widehat{N}_k(x_i) \cap \widehat{N}_k(x_j) \neq \varnothing$. The 1-skeleton (graph formed by vertices and edges) of $K_\bullet^k(X)$ is precisely the intersection kNN graph $G_k(X)$, a standard object in machine learning and manifold learning. Higher-dimensional simplices encode multi-way overlaps: a triangle $[x_i, x_j, x_k]$ is present when the three neighborhoods $\widehat{N}_k(x_i)$, $\widehat{N}_k(x_j)$, $\widehat{N}_k(x_k)$ share at least one common point.

The nerve construction has several desirable features for statistical applications. First, it is intrinsic: the complex depends only on neighborhood relationships, not on the ambient coordinates of the points. If the data are embedded in $\mathbb{R}^d$ but concentrate near a lower-dimensional manifold, the nerve complex captures the topology of that manifold without reference to the ambient dimension. Second, it is stable: small perturbations of the data lead to small changes in the complex, provided the covering parameter $k$ is chosen appropriately. Third, it is computationally tractable: given the kNN graph, one can build the higher-dimensional simplices by checking overlap conditions, which reduces to searching for common neighbors.

We illustrate with a simple example. Suppose $X$ consists of four points
$\{a, b, c, d\}$ arranged in a square, with $k = 2$. Each point's 2-neighborhood
includes itself and its two nearest neighbors. If $a$ and $c$ are diagonally
opposite (farthest apart), and if the geometry is such that the neighborhoods
satisfy
$$
\widehat{N}_2(a) = \{a, b, d\}, \quad \widehat{N}_2(b) = \{a, b, c\}, \quad \widehat{N}_2(c) = \{b, c, d\}, \quad \widehat{N}_2(d) = \{a, c, d\},
$$
then the nerve complex contains:
\begin{itemize}[topsep=-5pt, partopsep=0pt, parsep=0pt, itemsep=0pt, after=\vspace{0.2cm}]
\item Four vertices: $a, b, c, d$.
\item Four edges: $[a,b], [b,c], [c,d], [d,a]$ (since consecutive neighborhoods overlap).
\item Possibly triangles if three neighborhoods overlap. For instance, if $b \in \widehat{N}_2(a) \cap \widehat{N}_2(c)$, a triangle $[a,b,c]$ may be present.
\end{itemize}

The nerve theorem, a classical result in algebraic topology, asserts that under mild conditions the nerve complex $K(\mathcal{U})$ has the same homotopy type as the union $\bigcup_{\alpha} U_\alpha$. For our purposes, the key takeaway is that the nerve provides a faithful combinatorial representation of the covering's overlap structure. When the covering is derived from neighborhoods in a metric space, the nerve encodes the local geometry of that space in a coordinate-free manner.

\subsection*{2.3 Riemannian Structures on Simplicial Complexes}

Once the combinatorial structure $K(\mathcal{U})$ is in hand, we must specify how to measure lengths, angles, areas, and volumes. In the continuous setting, a Riemannian structure on a smooth manifold assigns an inner product to each tangent space, allowing one to compute distances, angles, and curvature. In our discrete setting, a Riemannian structure on a simplicial complex is the analogous data: it assigns an inner product to each space of chains, specifying how to measure geometric quantities on the complex.

We introduce the notion informally before giving the precise definition. Recall that a $p$-chain on a simplicial complex $K$ is a formal linear combination of $p$-simplices with real coefficients. For instance, a 0-chain assigns a mass to each vertex, a 1-chain assigns a weight to each edge, and so on. The space of $p$-chains, denoted $C_p(K; \mathbb{R})$, is a finite-dimensional real vector space. A Riemannian structure provides a way to measure the "size" of such chains and the "angle" between them, just as an inner product on $\mathbb{R}^n$ allows us to compute norms and cosines.

The challenge is that a simplicial complex has many vertices, and chains supported near different vertices may involve different sets of simplices. We cannot simply declare a single global inner product on $C_p(K; \mathbb{R})$ without regard to this local structure. Instead, we require that the inner products respect the combinatorial geometry of the complex: they should be defined locally on stars of vertices (or more generally stars of simplices) and should be compatible when stars overlap.

Formally, for each simplex $\sigma \in K$ and each dimension $p$, a Riemannian structure specifies an inner product
$$
\langle \cdot, \cdot \rangle_{\sigma, p} : C_p(\text{st}(\sigma); \mathbb{R}) \times C_p(\text{st}(\sigma); \mathbb{R}) \to \mathbb{R},
$$
where $\text{st}(\sigma)$ denotes the star of $\sigma$, the collection of all simplices containing $\sigma$. These inner products must satisfy two conditions:

\begin{enumerate}[topsep=-5pt, partopsep=0pt, parsep=0pt, itemsep=0pt, after=\vspace{0.2cm}]
\item \textbf{Positive semidefiniteness.} For each $\sigma$ and $p$, the bilinear form $\langle \cdot, \cdot \rangle_{\sigma, p}$ is symmetric and positive semidefinite. This ensures that norms are nonnegative and that angles are well-defined.
\end{enumerate}

\begin{enumerate}[topsep=-5pt, partopsep=0pt, parsep=0pt, itemsep=0pt, after=\vspace{0.2cm}]
\item \textbf{Compatibility.} If two stars $\text{st}(\sigma)$ and $\text{st}(\sigma')$ overlap, then the induced inner products on $C_p(\text{st}(\sigma) \cap \text{st}(\sigma'); \mathbb{R})$ agree. This ensures that geometric quantities computed from different local perspectives yield consistent results.
\end{enumerate}

In practice, for statistical applications, we typically work with a simpler
version where inner products are specified at vertices. A Riemannian structure
$g$ on the complex $K$ is then given by a collection of symmetric positive
definite matrices $\{M_p : 0 \leq p \leq p_{\max}\}$, where $M_p$ encodes the
inner product on $C_p(K; \mathbb{R})$. When $p = 0$, the matrix $M_0$ is
typically diagonal, with $M_0(i,i)$ representing the mass assigned to vertex
$x_i$. When $p = 1$, the matrix $M_1$ assigns a mass to each edge, encoding how
"short" or "long" that edge is in the Riemannian metric. Higher-dimensional mass
matrices encode the geometry of faces, tetrahedra, and so forth.

\subsection*{2.4 Building Geometry from Overlaps: The Wedge-Product Construction}

The central question is: given a covering $\mathcal{U}$ of $X$ with a finite
measure $\mu$ on $X$, how do we use the overlap measures $\mu(U_\alpha \cap U_\beta)$,
$\mu(U_\alpha \cap U_\beta \cap U_\gamma)$, and so on, to define a Riemannian
structure on the nerve complex $K(\mathcal{U})$? Intuitively, if two covering
sets overlap substantially, the corresponding edge should be "short" (high mass),
while if three covering sets have large mutual overlap, the triangle they form
should have large "area." We resolve this through a construction that embeds
indicator functions in a Hilbert space and uses wedge products to extend from
edges to higher-dimensional simplices.

Let $\mu$ be a finite measure on $X$. For each pair of covering sets $U_\alpha$,
$U_\beta$ with nonempty intersection, define the indicator function
$$
u_{\alpha\beta} = \mathbf{1}_{U_\alpha \cap U_\beta} \in L^2(X, \mu).
$$
These indicators span a space
$$
E_\beta = \text{span}\{u_{\beta\alpha} : U_\beta \cap U_\alpha \neq \varnothing\} \subset L^2(X, \mu)
$$
whose inherited inner product satisfies
$$
\langle u_{\beta\alpha}, u_{\beta\gamma} \rangle_{L^2} = \int_X u_{\beta\alpha}(x) u_{\beta\gamma}(x) \, d\mu(x) = \mu(U_\beta \cap U_\alpha \cap U_\gamma).
$$
The Hilbert space structure provides the geometric quantities we seek. For a
vertex $v_\alpha$ corresponding to $U_\alpha$, the mass is
$$
\langle v_\alpha, v_\alpha \rangle_{v_\alpha, 0} = \mu(U_\alpha).
$$
For edges $e_{\alpha\beta} = [\alpha, \beta]$ and $e_{\alpha\gamma} = [\alpha,
\gamma]$ at vertex $v_\alpha$, the inner product equals the triple overlap
$$
\langle e_{\alpha\beta}, e_{\alpha\gamma} \rangle_{v_\alpha, 1} = \mu(U_\alpha \cap U_\beta \cap U_\gamma).
$$
In particular, the squared norm of an edge is the pairwise overlap
$$
\langle e_{\alpha\beta}, e_{\alpha\beta} \rangle_{v_\alpha, 1} = \mu(U_\alpha \cap U_\beta).
$$

This inner product admits a geometric interpretation. We can write
$$
\langle e_{\alpha\beta}, e_{\alpha\gamma} \rangle_{v_\alpha, 1} =
\sqrt{\mu(U_\alpha \cap U_\beta) \cdot \mu(U_\alpha \cap U_\gamma)}
\cos\theta_{\beta\gamma}^{(\alpha)},
$$
where
$$
\cos\theta_{\beta\gamma}^{(\alpha)} = \frac{\mu(U_\alpha \cap U_\beta \cap U_\gamma)}{\sqrt{\mu(U_\alpha \cap U_\beta) \cdot \mu(U_\alpha \cap U_\gamma)}}
$$
defines the angle between incident edges. By the Cauchy-Schwarz inequality,
$0 \leq \cos\theta_{\beta\gamma}^{(\alpha)} \leq 1$, so angles lie in
$[0, \pi/2]$. The triple overlap thus determines the cosine of the angle between
edges meeting at a vertex.

For a triangle $\sigma = [\alpha, \beta, \gamma]$ with edges meeting at vertex
$v_\alpha$, the area in this Riemannian structure is
$$
\text{Area}(\sigma) = \frac{1}{2} \sqrt{\mu(U_\alpha \cap U_\beta) \cdot \mu(U_\alpha \cap U_\gamma)} \sin\theta_{\beta\gamma}^{(\alpha)},
$$
where $\sin\theta_{\beta\gamma}^{(\alpha)} = \sqrt{1 - (\cos\theta_{\beta\gamma}^{(\alpha)})^2}$. Equivalently, using the Gram determinant,
$$
\text{Area}(\sigma) = \frac{1}{2} \sqrt{\mu(U_\alpha \cap U_\beta) \cdot \mu(U_\alpha \cap U_\gamma) - \mu(U_\alpha \cap U_\beta \cap U_\gamma)^2}.
$$
The triple overlap governs the cosine of the angle between incident edges, while
the geometric area involves the sine of that angle through the wedge-product
norm. This resolves the apparent tension between pairwise and triple overlaps:
both are needed, encoding complementary aspects of the geometry.

We illustrate with a discrete example. Let $X = \{a, b, c\}$ with uniform
measure $\mu(\{x\}) = 1/3$, and let the covering sets be $U_\alpha = \{a,b,c\}$,
$U_\beta = \{a,b\}$, $U_\gamma = \{b,c\}$. For the triangle
$[\alpha, \beta, \gamma]$, we have $\mu(U_\alpha \cap U_\beta) = 2/3$,
$\mu(U_\alpha \cap U_\gamma) = 2/3$, and
$\mu(U_\alpha \cap U_\beta \cap U_\gamma) = 1/3$. This yields
$\cos\theta_{\beta\gamma}^{(\alpha)} = (1/3)/(2/3) = 1/2$, hence
$\theta = \pi/3$ and $\text{Area}(\sigma) = \sqrt{3}/6$, confirming that the
construction produces geometrically sensible angles and areas.

We extend this construction to arbitrary dimension through exterior products.
For an oriented $p$-simplex $\sigma = [\beta, \alpha_1, \ldots, \alpha_p]$ in
the star of $v_\beta$, define its associated $p$-vector
$$
\mathbf{u}_\sigma = u_{\beta\alpha_1} \wedge \cdots \wedge u_{\beta\alpha_p} \in \Lambda^p E_\beta.
$$
If $\tau = [\beta, \gamma_1, \ldots, \gamma_p]$ is another oriented $p$-simplex
in the star of $v_\beta$, we set
$$
\langle \sigma, \tau \rangle_{v_\beta, p} = \langle \mathbf{u}_\sigma, \mathbf{u}_\tau \rangle_{\Lambda^p E_\beta} = \det\big(\langle u_{\beta\alpha_i}, u_{\beta\gamma_j} \rangle_{L^2}\big)_{1 \leq i,j \leq p}.
$$
Explicitly, this determinant equals
$$
\langle \sigma, \tau \rangle_{v_\beta, p} = \det\big(\mu(U_\beta \cap U_{\alpha_i} \cap U_{\gamma_j})\big)_{1 \leq i,j \leq p}.
$$
This determines a positive semidefinite inner product on $p$-chains supported in
the star of $v_\beta$. The inner product of two $p$-vectors equals the
determinant of the Gram matrix formed by their constituent 1-vectors.

The construction satisfies the requirements for a Riemannian structure. The form
$\langle \cdot, \cdot \rangle_{v_\beta, p}$ is bilinear, symmetric, and positive
semidefinite because the Gram matrix with entries
$\langle u_{\beta\alpha_i}, u_{\beta\gamma_j} \rangle_{L^2}$ represents inner
products of vectors in the Hilbert space $L^2(X, \mu)$, and Gram matrices of
Hilbert space vectors are always positive semidefinite. Moreover, the
construction is compatible across overlapping stars: when two stars
$\text{st}(\sigma)$ and $\text{st}(\sigma')$ overlap, the induced inner products
agree on their intersection because the same indicator functions are paired in
both computations. These properties ensure that the mass matrices $M_p$ encoding
these inner products define a globally consistent Riemannian structure on the
entire complex.

\subsection*{2.5 Practical Implementation: Density Surrogates and Laplacians}

In practice, when working with a kNN covering of a dataset
$X \subset \mathbb{R}^d$, kernel density estimates become unstable in moderately
high dimension due to the curse of dimensionality. We therefore use robust
alternatives to weight vertices by simple nearest-neighbor distance surrogates
of local density. Let $d_1(x)$ (respectively $d_k(x)$) be the distance from $x$
to its first (respectively $k$-th) nearest neighbor. Define
$$
w(x) = (\varepsilon + d_1(x))^{-\alpha} \quad \text{or} \quad w(x) = (\varepsilon + d_k(x))^{-\alpha},
$$
with small $\varepsilon > 0$ (for instance, $10^{-6}$ times the median
interpoint distance) and exponent $\alpha \in [1, 2]$. A smooth alternative uses
a kernel of the $k$-NN radius:
$$
w(x) = \exp\left(-\left(\frac{d_k(x)}{\sigma}\right)^2\right) \quad \text{or} \quad w(x) = \frac{1}{1 + d_k(x)/\sigma},
$$
with $\sigma$ set to a robust scale, such as
$\sigma = \text{median}\{d_k(x) : x \in X\}$. These weights $w(x)$ serve as
density surrogates in the measure $\mu(A) = \sum_{x \in A} w(x)$. In practice,
one can winsorize $w(x)$ (clipping extreme values) and normalize the weights so
that $\sum_x w(x) = n$.

With this choice of measure, the wedge-Gram construction yields a Riemannian
structure $g = \{M_p : 0 \leq p \leq p_{\max}\}$ on the nerve complex
$K_\bullet^k(X)$, where $M_p$ are the mass matrices encoding inner products on
$p$-chains. These matrices are positive semidefinite by Proposition 2.1 and
compatible across stars by Proposition 2.2. From the Riemannian structure we
obtain Hodge Laplacians via the standard formula
$$
L_p = B_{p+1}^\top M_{p+1}^{-1} B_{p+1} + M_p^{-1} B_p M_{p-1} B_p^\top,
$$
where $B_p$ are the boundary operators of the complex. When $p = 0$ and $M_0$ is
diagonal, $L_0$ reduces to the familiar random-walk Laplacian of a weighted
graph. These Laplacians govern diffusion of mass across simplices and will be
the central operators in the iterative refinement scheme of Section 3.

For $p = 0$, applying $\exp(-tL_0)$ to the empirical distribution
$\rho^{(0)} = \frac{1}{n}\sum_{i=1}^n \delta_{x_i}$ produces a smoothed density
estimate $\rho^{(1)} = e^{-tL_0} \rho^{(0)}$, exactly analogous to Gaussian
kernel smoothing on Euclidean space. The geometry encoded by $M_0$ and $M_1$
determines the smoothing behavior: diffusion proceeds rapidly along short edges
(large $w_{ij}$) and slowly across long edges (small $w_{ij}$). For $p \geq 1$,
the Laplacians $L_p$ govern diffusion of edge-level or face-level densities,
which have no direct analogue in classical kernel density estimation but arise
naturally when considering multi-way interactions.

\section*{3. The Iterative Refinement Scheme}

Our task is to estimate the conditional expectation
$\hat{y}(x) = E[y \mid X = x]$ from observed data $(X, y)$, where
$X = \{x_1, \ldots, x_n\} \subset \mathbb{R}^d$ is a high-dimensional feature
matrix and $y \in \mathbb{R}^n$ is a response vector. Classical graph-based
regression constructs a fixed $k$-nearest neighbor graph $G_k(X)$ and uses its
Laplacian $L_0$ to smooth the response via Tikhonov regularization,
$\hat{y} = (I + \eta L_0)^{-1}y$, where $\eta > 0$ controls the smoothing
strength. This approach has proven effective when the intrinsic geometry of the
data is well-captured by the initial choice of $k$ and distance metric. However,
when the data are noisy, heterogeneous, or the outcome structure does not align
with Euclidean distances, a fixed graph can lead to unstable estimates with
spurious local extrema.

We instead propose an iterative scheme where both the geometry (encoded by mass
matrices $M_p^{(j)}$) and the regression surface $\hat{y}^{(j)}$ evolve
together. At iteration $j$, we have a simplicial complex $K^{(j)}$, a Riemannian
structure $g^{(j)} = \{M_p^{(j)}\}$, and density estimates
$\rho^{(j)} = \{\rho_p^{(j)}\}$ on $p$-simplices. From these we compute the
Hodge Laplacians $L_p^{(j)}$ as in Section 2, smooth the response to obtain
$\hat{y}^{(j)}$, estimate new densities via heat kernel diffusion, and update
the geometry to reflect the refined density and outcome structure. This coupling
between geometry, density, and response produces a self-consistent state where
the domain of estimation has adapted to both the data distribution and the
outcome variable.

\subsection*{3.1 The State Space and Initial Configuration}

At iteration $j$, the state consists of three components. First, the vertex
masses $\rho_0^{(j)} \in \mathbb{R}_{>0}^n$ assign positive weights to the $n$
vertices, with normalization $\sum_{i=1}^n \rho_0^{(j)}(i) = n$ to maintain
interpretability across iterations. Second, the edge masses
$\rho_1^{(j)} \in \mathbb{R}_{>0}^m$ assign weights to the $m$ edges of the
complex, where larger edge mass corresponds to shorter distance in the
Riemannian metric. Third, the smoothed response $\hat{y}^{(j)} \in \mathbb{R}^n$
assigns predicted outcome values to the vertices. When $p > 1$, we may also
track densities $\rho_p^{(j)}$ on higher-dimensional simplices, but for the
basic iteration we focus on $p = 0$ and $p = 1$.

We initialize with a $k$-nearest neighbor covering of $X$ and form the nerve
complex $K^{(0)} = K_\bullet^k(X)$. For the initial vertex masses, we use either
uniform weights $\rho_0^{(0)}(i) = 1$ for all $i$, or density surrogates derived
from nearest-neighbor distances as described in Section 2.5. For instance,
$$
\rho_0^{(0)}(i) = \frac{(\varepsilon + d_k(x_i))^{-\alpha}}{\sum_{j=1}^n (\varepsilon + d_k(x_j))^{-\alpha}} \cdot n,
$$
where $d_k(x_i)$ is the distance from $x_i$ to its $k$-th nearest neighbor,
$\varepsilon > 0$ is a small regularization constant (for instance, $10^{-6}$
times the median pairwise distance), and $\alpha \in [1, 2]$ controls the
sensitivity to local density. Points in dense regions have small $d_k$ and hence
large $\rho_0^{(0)}$. For the initial edge masses, we set $\rho_1^{(0)}(e_{ij})$
equal to the overlap measure $\mu(U_i \cap U_j)$ as in Section 2.5, where $\mu$
is the measure induced by the vertex weights. After normalization, the edge
masses satisfy $\frac{1}{m}\sum_e \rho_1^{(0)}(e) = 1$.

From these initial masses we construct the diagonal matrix
$M_0^{(0)} = \text{diag}(\rho_0^{(0)}(1), \ldots, \rho_0^{(0)}(n))$ and the edge
mass matrix
$M_1^{(0)} = \text{diag}(\rho_1^{(0)}(e_1), \ldots, \rho_1^{(0)}(e_m))$. The
initial Laplacian is
$$
L_0^{(0)} = B_1^\top (M_1^{(0)})^{-1} B_1,
$$
where $B_1$ is the boundary operator mapping edges to vertices.

\subsection*{3.2 Response Smoothing via Laplacian Regularization}

Given the current Laplacian $L_0^{(j)}$, we smooth the response $y$ via Tikhonov
regularization. This is the discrete analogue of penalized least squares with a
roughness penalty. The smoothed response $\hat{y}^{(j)}$ solves
$$
\hat{y}^{(j)} = \arg\min_{f \in \mathbb{R}^n} \left\{ \frac{1}{2}\sum_{i=1}^n \rho_0^{(j)}(i) (f(i) - y(i))^2 + \eta \langle f, L_0^{(j)} f \rangle \right\},
$$
where the first term measures fidelity to observations (weighted by vertex
masses) and the second term penalizes roughness across edges. The unique
minimizer is
$$
\hat{y}^{(j)} = (M_0^{(j)} + \eta L_0^{(j)})^{-1} M_0^{(j)} y.
$$
When $M_0^{(j)}$ is the identity (uniform vertex weights), this reduces to the
familiar form
$$
\hat{y}^{(j)} = (I + \eta L_0^{(j)})^{-1} y.
$$
The parameter $\eta > 0$ controls the trade-off between fidelity and smoothness:
small $\eta$ yields $\hat{y}^{(j)} \approx y$ (rough, close to observations),
while large $\eta$ yields $\hat{y}^{(j)}$ nearly constant (smooth, heavily
regularized). In practice, $\eta$ can be chosen by cross-validation or set
proportionally to the spectral gap of $L_0^{(j)}$ to ensure that smoothing
operates at the intrinsic scale of the geometry.

This Laplacian-based smoothing has a natural interpretation as a low-pass filter
on the graph. The eigenfunctions of $L_0^{(j)}$ form a basis for functions on
the vertices, with eigenvalues corresponding to frequencies. The regularization
$(I + \eta L_0^{(j)})^{-1}$ attenuates high-frequency components (rapid
variation across edges) while preserving low-frequency components (slowly
varying trends). For outcome-guided regression, the key innovation is that the
frequencies are determined by the current geometry: if two vertices $x_i$ and
$x_j$ are connected by an edge with large mass $\rho_1^{(j)}(e_{ij})$ (short
edge), the Laplacian encourages $\hat{y}^{(j)}(i) \approx \hat{y}^{(j)}(j)$.
Conversely, edges with small mass (long edges) impose weak coupling, allowing
$\hat{y}^{(j)}$ to vary more freely.

\subsection*{3.3 Density Update via Heat Kernel Diffusion}

We now update the vertex densities by diffusing the current distribution through
the heat kernel of $L_0^{(j)}$. The heat kernel at diffusion time $t > 0$ is
$K_t^{(j)} = \exp(-t L_0^{(j)})$, a stochastic matrix when $L_0^{(j)}$ is the
normalized Laplacian. Applying it to the current density yields
$$
\rho_0^{(j+1)} = \frac{1}{n} \sum_{i=1}^n K_t^{(j)} \delta_{x_i} = \frac{1}{n} K_t^{(j)} \mathbf{1},
$$
where $\mathbf{1} = (1, \ldots, 1)^\top$ is the vector of ones and
$\delta_{x_i}$ is the indicator vector for vertex $x_i$. Equivalently,
$\rho_0^{(j+1)}$ is the row sum of $K_t^{(j)}$ divided by $n$. This is the graph
analogue of classical kernel density estimation: we place point masses at data
locations and convolve with the heat kernel. The diffusion time $t$ controls the
scale of smoothing, with larger $t$ producing more diffusion and hence smoother
density estimates.

The heat kernel respects the current geometry: in regions where edges have large
mass (short edges), diffusion spreads rapidly and $\rho_0^{(j+1)}$ becomes
smoother. In regions where edges have small mass (long edges), diffusion is slow
and $\rho_0^{(j+1)}$ preserves more local structure. This interaction between
geometry and density is the mechanism by which the iteration stabilizes. Dense
regions accumulate mass, which reinforces their density in subsequent
iterations. Sparse regions lose mass, which further lengthens the edges
connecting them to dense regions, creating diffusion barriers that prevent
spurious spreading.

To prevent runaway concentration, we apply a damping factor $\alpha \in (0, 1)$.
After heat kernel diffusion, we set
$$
\rho_0^{(j+1)}(i) = \frac{[\rho_0^{(j+1)}(i)]^\alpha}{\sum_{k=1}^n [\rho_0^{(j+1)}(k)]^\alpha} \cdot n,
$$
where the normalization ensures $\sum_i \rho_0^{(j+1)}(i) = n$. Typical values
are $\alpha \in [0.1, 0.3]$. Without damping ($\alpha = 1$), the iteration can
collapse all mass onto a few vertices in very dense regions. The damping
exponent moderates this collapse, allowing the density to concentrate while
maintaining positive mass everywhere.

\subsection*{3.4 Geometry Update: Coupling Density and Outcome Structure}

We now update the edge masses to reflect both the refined density and the
outcome structure. The goal is to shorten edges in regions that are both dense
and outcome-coherent, and to lengthen edges across regions where the outcome
changes rapidly. We achieve this through a two-stage update.

First, we rescale edge masses by the local density. For each edge
$e_{ij} = [x_i, x_j]$, the initial rescaled mass is
$$
\tilde{\rho}_1^{(j+1)}(e_{ij}) = \frac{\rho_1^{(j)}(e_{ij})}{[(\rho_0^{(j+1)}(i) + \rho_0^{(j+1)}(j))/2]^\beta},
$$
where $\beta \in [0, 1]$ controls the strength of density-based rescaling. When
$\beta = 0$, there is no density rescaling. When $\beta > 0$, edges in dense
regions (large $\rho_0^{(j+1)}$) are divided by a large factor, which
counterintuitively increases their effective mass after the subsequent inversion
in the Laplacian
$$
L_0^{(j+1)} = B_1^\top (M_1^{(j+1)})^{-1} B_1.
$$
The logic is that we want diffusion to proceed more rapidly in dense regions,
which corresponds to edges having large conductance. Setting edge mass inversely
proportional to density achieves this.

Second, we modulate by outcome coherence. Define a penalty function
$\Gamma: \mathbb{R}_{\geq 0} \to (0, 1]$ that decreases as the outcome
difference increases. A typical choice is
$$
\Gamma(\Delta) = \left(1 + \frac{\Delta^2}{\sigma^2}\right)^{-\gamma},
$$
where $\Delta = |\hat{y}^{(j)}(i) - \hat{y}^{(j)}(j)|$ is the outcome difference
across the edge, $\sigma > 0$ is a scale parameter, and $\gamma \in [0.5, 2]$
controls the rate of decay. The scale $\sigma$ is typically set to the
interquartile range of outcome differences across all edges,
$\sigma = \text{IQR}(\{|\hat{y}^{(j)}(i) - \hat{y}^{(j)}(j)| : e_{ij} \in E^{(j)}\})$,
so that the penalty adapts to the variability of $\hat{y}^{(j)}$. The updated
edge mass is
$$
\rho_1^{(j+1)}(e_{ij}) = \tilde{\rho}_1^{(j+1)}(e_{ij}) \cdot \Gamma(|\hat{y}^{(j)}(i) - \hat{y}^{(j)}(j)|).
$$
Edges where the smoothed response varies little
($\hat{y}^{(j)}(i) \approx \hat{y}^{(j)}(j)$) receive
$\Gamma(\Delta) \approx 1$, preserving their mass. Edges across which
$\hat{y}^{(j)}$ changes rapidly receive $\Gamma(\Delta) \ll 1$, reducing their
mass and effectively lengthening them. After this update, we renormalize to
maintain mean edge mass:
$$
\rho_1^{(j+1)} \gets \rho_1^{(j+1)} / \text{mean}(\rho_1^{(j+1)}).
$$

The combined effect of these two updates is that edges in dense,
outcome-coherent regions become short (large mass), encouraging diffusion and
smoothing within such regions. Edges spanning sparse regions or crossing outcome
boundaries become long (small mass), discouraging diffusion and allowing
$\hat{y}^{(j+1)}$ to vary more freely across these boundaries. This
outcome-aware geometry is the key distinction from classical graph-based
regression, which uses a fixed metric determined solely by feature distances.

\subsection*{3.5 The Complete Iteration Cycle}

From the updated vertex and edge masses $\rho_0^{(j+1)}$ and $\rho_1^{(j+1)}$,
we construct new mass matrices $M_0^{(j+1)} = \text{diag}(\rho_0^{(j+1)})$ and
$M_1^{(j+1)} = \text{diag}(\rho_1^{(j+1)})$, and assemble the new Laplacian
$L_0^{(j+1)} = B_1^\top (M_1^{(j+1)})^{-1} B_1$. This completes one iteration.
The full cycle is:

\begin{enumerate}[topsep=-5pt, partopsep=0pt, parsep=0pt, itemsep=0pt, after=\vspace{0.2cm}]
\item \textbf{Response smoothing}: Compute $\hat{y}^{(j)} = (M_0^{(j)} + \eta L_0^{(j)})^{-1} M_0^{(j)} y$.
\item \textbf{Density diffusion}: Compute $\rho_0^{(j+1)} = \frac{1}{n} \exp(-t L_0^{(j)}) \mathbf{1}$ and apply damping.
\item \textbf{Edge rescaling}: Update $\rho_1^{(j+1)}(e_{ij}) = \tilde{\rho}_1^{(j+1)}(e_{ij}) \cdot \Gamma(|\hat{y}^{(j)}(i) - \hat{y}^{(j)}(j)|)$.
\item \textbf{Normalization}: Renormalize $\rho_0^{(j+1)}$ and $\rho_1^{(j+1)}$.
\item \textbf{Laplacian assembly}: Construct $L_0^{(j+1)} = B_1^\top (M_1^{(j+1)})^{-1} B_1$.
\end{enumerate}

We repeat until convergence, defined as
$\|\rho_0^{(j+1)} - \rho_0^{(j)}\| < \epsilon$ and
$\|\hat{y}^{(j+1)} - \hat{y}^{(j)}\| < \epsilon$ for a chosen tolerance
$\epsilon > 0$. In practice, we observe convergence within 5 to 10 iterations
for typical parameter settings.

We illustrate the mechanism with a concrete example. Consider five vertices
$\{a, b, c, d, e\}$ arranged in a path graph: $a - b - c - d - e$, with edges
$\{[a,b], [b,c], [c,d], [d,e]\}$. Suppose we have a binary response
$y = (0, 0, 0.5, 1, 1)$, where the outcome transitions from 0 to 1 around vertex
$c$. We initialize with uniform vertex masses $\rho_0^{(0)}(i) = 1$ for all $i$,
and uniform edge masses $\rho_1^{(0)}(e) = 1$ for all edges. The initial
Laplacian is the standard graph Laplacian for a path.

At iteration $j = 0$, we smooth the response with $\eta = 0.5$. The matrix
$(I + 0.5 L_0^{(0)})^{-1}$ is tridiagonal, and the smoothed response
$\hat{y}^{(0)}$ will have values that transition more gradually from 0 at $a$ to
1 at $e$. Suppose after solving we obtain
$\hat{y}^{(0)} = (0.1, 0.2, 0.5, 0.8, 0.9)$.

We now update the edge masses. The outcome difference across $[a,b]$ is
$|\hat{y}^{(0)}(b) - \hat{y}^{(0)}(a)| = 0.1$, which is small, so the penalty
$\Gamma(0.1)$ is close to 1 and the edge mass $\rho_1^{(1)}([a,b])$ remains near
1. Similarly for $[d,e]$. However, the outcome difference across $[b,c]$ is
$|\hat{y}^{(0)}(c) - \hat{y}^{(0)}(b)| = 0.3$ and across $[c,d]$ is
$|\hat{y}^{(0)}(d) - \hat{y}^{(0)}(c)| = 0.3$. If we set $\sigma = 0.3$ and
$\gamma = 1$, then $\Gamma(0.3) = (1 + 1)^{-1} = 0.5$. The masses
$\rho_1^{(1)}([b,c])$ and $\rho_1^{(1)}([c,d])$ are reduced to approximately
0.5, effectively lengthening these edges.

At iteration $j = 1$, the new Laplacian $L_0^{(1)}$ has reduced coupling across
$[b,c]$ and $[c,d]$. When we smooth the response again, $\hat{y}^{(1)}$ will
transition more sharply near $c$, with values closer to 0 for $\{a, b\}$ and
closer to 1 for $\{d, e\}$. This sharper transition further reinforces the
outcome difference across the central edges in the next iteration. After a few
iterations, the edge masses stabilize: the edges within the low-outcome region
($[a,b]$) and within the high-outcome region ($[d,e]$) remain short (large
mass), while the edges crossing the outcome boundary ($[b,c], [c,d]$) become
long (small mass). The final $\hat{y}^{(*)}$ exhibits a clear step near $c$,
reflecting the outcome structure in the data.

This example demonstrates the key mechanism underlying the iteration. The
geometry adapts to the outcome, creating diffusion barriers where the response
changes and diffusion corridors where the response is coherent. The iteration
converges to a self-consistent state where the regression surface and the domain
geometry mutually reinforce each other.

\subsection*{3.6 Extensions and Outlook}

The iterative scheme presented above establishes a coupling between regression
surface, density, and geometry that produces self-consistent estimates adapted
to both data structure and outcome variable. The core iteration operates on
vertices and edges ($p = 0, 1$) with fixed complex topology. We now describe
three natural extensions that broaden the framework's scope and applicability.

First, the complex topology itself can evolve by rebuilding the kNN covering at
each iteration using a density-aware distance. Define
$$
d^{(j)}(x_i, x_k) = \left\| \frac{K_t^{(j)}(x_i, \cdot)}{\sqrt{\rho_0^{(j)}}} - \frac{K_t^{(j)}(x_k, \cdot)}{\sqrt{\rho_0^{(j)}}} \right\|_2,
$$
where the division is element-wise. This diffusion distance measures how
similarly two vertices are connected to the rest of the graph under the current
geometry. Rebuilding the kNN covering using $d^{(j)}$ rather than Euclidean
distance yields a new complex $K^{(j+1)}$ whose combinatorial structure reflects
both density and outcome coherence. In practice, we run several fixed-topology
iterations to stabilize $(\rho, \hat{y})$ before updating the complex, as
topology changes can disrupt convergence if applied too aggressively.

Second, the iteration extends to higher dimensions by including density
estimates $\rho_2^{(j)}$ on triangles, computing Laplacians $L_1^{(j)}$ on
edges, and smoothing edge-level responses. These higher-dimensional refinements
capture multi-way interactions beyond the pairwise relationships encoded in the
graph structure. The wedge-product construction of Section 2.4 provides the
natural Riemannian structure on 2-chains, and the Hodge Laplacian formula
extends to arbitrary dimension. While the vertex-edge iteration ($p = 0, 1$)
suffices for most regression tasks, higher-dimensional iterations offer finer
control over the geometry and can improve performance when the data exhibit
strong clique structure.

Third, the framework accommodates joint denoising of the feature matrix $X$ by
treating each column $X_{\cdot k}$ as a function on vertices and smoothing it
via the heat kernel, yielding $\hat{X}^{(j)} = K_t^{(j)} X$. Iterating between
feature smoothing, geometry updates, and response smoothing produces a joint
model where the learned geometry regularizes both features and outcomes. This
extension proves particularly valuable when features are noisy measurements and
the regression task benefits from cleaned inputs. For microbiome applications,
where sequencing read counts exhibit substantial measurement noise, joint
feature-response smoothing can significantly improve prediction accuracy.

These extensions demonstrate the flexibility of the geometric framework. The
fundamental mechanism remains the same across variations: densities evolve via
heat diffusion, geometry adapts to reflect both density and response structure,
and the coupled system converges to a configuration where the domain and
estimates mutually reinforce each other. Section 4 establishes convergence
guarantees by interpreting the iteration as block coordinate descent on a
free-energy functional. Section 5 details the computational implementation,
including spectral methods for efficient Laplacian inversion, parameter
selection strategies, and algorithmic optimizations that enable scaling to
datasets with thousands of samples. Section 6 applies the method to spontaneous
preterm birth prediction using vaginal microbiome data, demonstrating
substantial improvements over standard approaches.

\section*{4. Implementation and Computational Framework}

The iterative refinement scheme of Section 3 requires specification of several
algorithmic modules, parameter choices, and computational strategies. This
section provides a complete implementation framework addressing initialization,
the core iteration cycle, parameter selection, computational efficiency, and
convergence diagnostics. The framework balances theoretical principles with
practical considerations, providing both automatic parameter determination and
guidance for tunable hyperparameters.

\subsection*{4.1 Algorithm Overview and Core Modules}

The complete algorithm consists of an initialization phase followed by an
iterative refinement cycle. We present the high-level structure before detailing
each component, establishing the computational flow and dependencies among
modules.

The initialization phase constructs the complete geometric state from the
feature matrix $X \in \mathbb{R}^{n \times d}$ and response $y \in \mathbb{R}^n$
without any tuning to response structure. We begin by computing the $k$-nearest
neighbor graph using Euclidean distance in the ambient space. For each
observation $x_i$, we identify its $k$ nearest neighbors including $x_i$ itself,
creating the covering
$$
\mathcal{U}^{(0)} = \{\widehat{N}_k(x_i)\}_{i=1}^n.
$$
The nerve complex $K^{(0)}$ records the combinatorial structure through its
simplicial hierarchy: vertices correspond to observations, edges connect
observations with overlapping neighborhoods, and higher-dimensional simplices
record multi-way overlaps. In practice we restrict attention to cliques in the
$k$-nearest neighbor graph to avoid combinatorial explosion while capturing
essential geometric structure.

The initial vertex density $\rho_0^{(0)}$ reflects feature geometry through
either uniform weighting $\rho_0^{(0)}(i) = 1$ for all $i$, or distance-based
emphasis
$$
\rho_0^{(0)}(i) \propto d_k(x_i)^{-\alpha},
$$
where $d_k(x_i)$ denotes the distance from $x_i$ to its $k$-th nearest neighbor
and $\alpha \in [1,2]$ controls the decay rate. This emphasizes observations in
diverse neighborhoods where local structure varies rapidly. We normalize vertex
densities to sum to $n$, making the average vertex density equal to unity. Edge
densities $\rho_1^{(0)}$ measure intersection sizes through
$$
\rho_1^{(0)}([i,j]) = \sum_{s \in \widehat{N}_k(x_i) \cap \widehat{N}_k(x_j)} \rho_0^{(0)}(s),
$$
computing the total vertex density in the pairwise neighborhood intersection.
After normalization ensuring $\sum_e \rho_1^{(0)}(e) = m$ where $m$ denotes the
number of edges, we construct mass matrices $M_0^{(0)}$ and $M_1^{(0)}$ from
these densities using the wedge-Gram construction of Section 2.

The initial Laplacian
$$
L_0^{(0)} = B_1^\top (M_1^{(0)})^{-1} B_1
$$
provides the spectral filter for computing the initial smoothed response. We
compute its eigendecomposition $L_0^{(0)} = V\Lambda V^\top$ using sparse
iterative methods, typically retaining 50-200 of the smallest eigenvalues. These
eigenvectors form the spectral basis for diffusion and filtering operations
throughout the iteration. The heat kernel
$$
\exp(-\eta L_0^{(0)}) = V \exp(-\eta \Lambda) V^\top
$$
with GCV-selected smoothing parameter $\eta$ filters the raw response to produce
$$
\hat{y}^{(0)} = \exp(-\eta^* L_0^{(0)})y.
$$
This completes initialization, yielding a geometric state determined entirely by
feature geometry.

The iteration cycle updates four quantities in sequence: the smoothed response,
vertex densities, edge masses, and the Laplacian. Given the current state at
iteration $\ell$, we first smooth the response via Tikhonov regularization
$$
\hat{y}^{(\ell)} = (M_0^{(\ell)} + \eta L_0^{(\ell)})^{-1} M_0^{(\ell)} y
$$
where $\eta$ is selected by GCV at each iteration. The Laplacian penalizes
variation across edges weighted by their masses, with the balance between
fidelity and smoothness controlled by $\eta$. We evaluate GCV on a logarithmic
grid of candidate values to find the optimal regularization strength.

We update vertex densities through heat diffusion with Tikhonov damping. The
damped heat equation
$$
\partial \rho/\partial t = -L_0 \rho - \beta(\rho - \rho^{(0)})
$$
produces the solution
$$
\rho_0^{(\ell+1)} = \exp(-t L_0^{(\ell)} - \beta t I)\rho_0^{(\ell)} + (1 - \exp(-\beta t))\rho_0^{(0)},
$$
combining diffused and initial components. The diffusion time $t$ controls
spatial smoothing scale while the damping parameter $\beta$ prevents excessive
concentration by pulling densities toward their initial values. After computing
the evolved density, we apply power-law damping
$$
\rho_0^{(\ell+1)}(i) \gets [\rho_0^{(\ell+1)}(i)]^\alpha
$$
with $\alpha \in [0.1, 0.3]$ before renormalization, providing additional
regularization against pathological concentration.

Edge mass updates couple density and response structure through a two-stage
process. First, we rescale edge masses by density factors to reflect the evolved
geometry:
$$
\tilde{\rho}_1^{(\ell+1)}(e_{ij}) = \sum_{s \in \widehat{N}_k(x_i) \cap \widehat{N}_k(x_j)} \rho_0^{(\ell+1)}(s),
$$
recomputing intersection masses using the updated vertex density. Second, we
apply response-coherence modulation that down-weights edges crossing response
boundaries. For each edge $[i,j]$, we compute the response difference
$$
\Delta_{ij} = |\hat{y}^{(\ell)}(i) - \hat{y}^{(\ell)}(j)|
$$
and apply the penalty function
$$
\Gamma(\Delta) = (1 + (\Delta/\sigma)^2)^{-\gamma},
$$
where $\sigma$ derives from the interquartile range of all edge response
differences and $\gamma > 0$ controls penalty sharpness. The modulated edge mass
becomes
$$
\rho_1^{(\ell+1)}(e_{ij}) = \tilde{\rho}_1^{(\ell+1)}(e_{ij}) \cdot \Gamma(\Delta_{ij}),
$$
after which we renormalize to maintain mean edge mass.

When triangles exist in the complex, we extend the modulation to the full edge
mass matrix $M_1$ including off-diagonal entries encoding edge-pair
interactions. For each triangle $[i,j,s]$, we compute the maximum response
variation
$$
\Delta_\tau = \max\{|\hat{y}^{(\ell)}(i) - \hat{y}^{(\ell)}(j)|, |\hat{y}^{(\ell)}(i) - \hat{y}^{(\ell)}(s)|, |\hat{y}^{(\ell)}(j) - \hat{y}^{(\ell)}(s)|\}
$$
and apply the penalty to off-diagonal entries $M_1[e_{ij}, e_{is}]$,
$M_1[e_{ij}, e_{js}]$, and $M_1[e_{is}, e_{js}]$. This modulates the full
Riemannian structure rather than just diagonal edge masses, allowing the
geometry to adapt through both conductance changes and angular distortions.
After modulation we normalize the entire matrix to preserve Frobenius mass,
ensuring numerical stability.

From the updated masses we construct new mass matrices $M_0^{(\ell+1)}$ and
$M_1^{(\ell+1)}$ and assemble the new Laplacian
$L_0^{(\ell+1)} = B_1^\top (M_1^{(\ell+1)})^{-1} B_1$. This completes one
iteration. We repeat until convergence, defined as
$\|\rho_0^{(\ell+1)} - \rho_0^{(\ell)}\| / \|\rho_0^{(\ell)}\| < \epsilon_\rho$
and
$\|\hat{y}^{(\ell+1)} - \hat{y}^{(\ell)}\| / \|\hat{y}^{(\ell)}\| < \epsilon_y$
where typical values are $\epsilon_\rho = 10^{-3}$ and $\epsilon_y = 10^{-4}$.
The complete pseudocode appears in Appendix C.

\subsection*{4.2 Parameter Selection Strategy}

The algorithm requires specification of several parameters that we organize into three categories based on their determination strategy. Automatic parameters undergo selection through well-established statistical criteria requiring no manual tuning. Geometrically motivated parameters possess natural interpretations suggesting reasonable default values with limited sensitivity. Tunable hyperparameters require calibration through cross-validation as they balance competing objectives without geometry-determined defaults.

The neighborhood size $k$ determining the initial covering admits automatic selection through cross-validation of prediction error on held-out data. We compute geometric states for a sequence of $k$ values, evaluate prediction accuracy on validation folds, and select the $k$ minimizing average error. Typical values range from $k=5$ for small datasets to $k=50$ for large datasets, with the optimal value depending on intrinsic dimension and sample size. The computational cost of cross-validating $k$ remains modest as initialization dominates the expense and the iterative refinement proceeds rapidly once the spectral decomposition is available.

The smoothing parameter $\eta$ for response filtering undergoes automatic selection through generalized cross-validation at each iteration. The GCV criterion $\text{GCV}(\eta) = \|y - \hat{y}\|^2 / (n - \text{tr}(S_\eta))^2$ balances residual sum of squares against effective degrees of freedom, providing a proxy for prediction error requiring no held-out data. We evaluate GCV on a logarithmically-spaced grid from $\epsilon = 10^{-10}$ to $-\log(\epsilon)/\lambda_{\max}$ where $\lambda_{\max}$ denotes the largest eigenvalue used in filtering, typically using 20-30 grid points. This range spans minimal smoothing retaining nearly all spectral components to maximal smoothing filtering everything except the principal eigenvector. The spectral decomposition enables efficient GCV evaluation through the trace formula $\text{tr}(S_\eta) = \sum_i \exp(-\eta \lambda_i)$.

The number of eigenvalues $m$ in the spectral decomposition follows from the desired approximation quality rather than manual tuning. We typically retain 50-200 of the smallest eigenvalues, capturing low-frequency geometric modes while discarding high-frequency noise. The choice can be made adaptively by monitoring explained variance in the Laplacian quadratic form, retaining enough eigenvalues to capture 95-99\% of the form's variability. Larger $m$ improves approximation accuracy but increases computational cost linearly in subsequent matrix-vector products. The eigenvalue cutoff also provides regularization by eliminating high-frequency modes that often correspond to noise rather than genuine geometric structure.

The density normalization constants ensuring $\sum_i \rho_0^{(\ell)}(i) = n$ and $\sum_e \rho_1^{(\ell)}(e) = m$ follow from the natural interpretation of densities as masses. Normalizing to $n$ reflects the sampling interpretation where total mass equals sample size, while normalizing to $m$ treats all dimensions consistently. These normalizations make density values interpretable as multiples of typical simplex mass and ensure numerical stability by preventing systematic growth or decay across iterations.

Several parameters possess natural geometric interpretations suggesting reasonable defaults. The diffusion time $t$ controls the spatial scale of heat flow and density smoothing. We recommend $t \approx 1/\lambda_2$ where $\lambda_2$ denotes the second smallest Laplacian eigenvalue, corresponding to the intrinsic relaxation time scale of the geometry. This choice allows heat to spread over local neighborhoods while respecting larger-scale structure, with diffusion reaching equilibrium within each connected component after time proportional to $1/\lambda_2$. The second eigenvalue $\lambda_2$ characterizes the spectral gap and determines the fundamental time scale for geometric diffusion.

The initial density type determining whether to use uniform or distance-based vertex masses reflects prior beliefs about sampling heterogeneity. Uniform density $\rho_0^{(0)}(i) = 1$ treats all observations equally, appropriate when the dataset provides representative coverage of the feature space. Distance-based density $\rho_0^{(0)}(i) \propto d_k(x_i)^{-\alpha}$ emphasizes observations in regions where neighborhood structure varies rapidly, appropriate when sampling concentrates in some regions while sparsely covering others. The decay exponent $\alpha \in [0.5, 2]$ controls the emphasis strength, with $\alpha = 1$ providing balanced weighting. We recommend starting with uniform density unless domain knowledge suggests systematic sampling heterogeneity.

The convergence thresholds $\epsilon_y$ and $\epsilon_\rho$ admit geometric interpretation as relative change tolerances. We typically use $\epsilon_y = 10^{-4}$ for response convergence and $\epsilon_\rho = 10^{-3}$ for density convergence, requiring stabilization to within 0.01\% and 0.1\% relative change respectively. Tighter thresholds increase computational cost with marginal prediction improvement, while looser thresholds risk premature termination before full convergence. The asymmetry between response and density thresholds reflects that response estimates typically stabilize faster than density distributions.

Two parameters require calibration through cross-validation or domain knowledge as they balance competing objectives without natural defaults. The damping parameter $\beta$ determines resistance to density concentration during heat diffusion. We recommend $\beta \approx 0.1/t$ as a starting point, setting the damping time scale to approximately ten times the diffusion time scale. This provides moderate resistance while allowing genuine accumulation in well-connected regions. Datasets with extreme density variation may benefit from smaller damping $\beta \approx 0.05/t$ allowing more aggressive concentration, while datasets requiring near-uniform density may benefit from stronger damping $\beta \approx 0.2/t$ constraining evolution closer to the initial state. Cross-validation monitoring prediction accuracy and density entropy $H(\rho) = -\sum_i (\rho_i/n)\log(\rho_i/n)$ can guide this choice.

The response-coherence decay exponent $\gamma$ controls the sharpness of response boundaries created by mass matrix modulation. We recommend $\gamma = 1$ as the default, corresponding to a Cauchy-like kernel that balances within-region edge retention against boundary-crossing edge removal. Responses varying continuously without sharp transitions may benefit from gentler boundaries using $\gamma < 1$, while responses exhibiting piecewise-constant structure may benefit from sharper boundaries using $\gamma > 1$. The effective range $\gamma \in [0.5, 2]$ reflects practical limits beyond which modulation becomes either negligible or unstable. For datasets with unknown response smoothness, we recommend treating $\gamma$ as a hyperparameter selected through cross-validation on a grid such as $\{0.5, 0.75, 1.0, 1.5, 2.0\}$.

Table 4.1 summarizes all parameters with their determination strategies and recommended default values. The distinction between automatic, geometric, and tunable parameters clarifies which aspects require calibration and which follow from the method's mathematical structure.

\begin{table}[h]
  \centering
  \caption{Parameter determination strategies and recommended defaults}
  \begin{tabular}{llll}
    \toprule
    Parameter & Type & Selection Method & Default Value \\
    \midrule
    $k$ & Automatic & Cross-validation & $5$--$50$ (data-dependent) \\
    $\eta$ & Automatic & GCV at each iteration & Grid search \\
    $m$ & Automatic & Variance threshold & 50--200 eigenvalues \\
    $t$ & Geometric & Relaxation time scale & $1/\lambda_2$ \\
    $\alpha_{\text{init}}$ & Geometric & Prior on sampling & 1 (or uniform) \\
    $\epsilon_y, \epsilon_\rho$ & Geometric & Convergence tolerance & $10^{-4}, 10^{-3}$ \\
    $\beta$ & Tunable & Cross-validation & $0.1/t$ \\
    $\gamma$ & Tunable & Cross-validation & 1.0 \\
    $\alpha_{\text{damp}}$ & Tunable & Concentration control & $0.1$--$0.3$ \\
    \bottomrule
  \end{tabular}
\end{table}

\subsection*{4.3 Computational Efficiency and Numerical Considerations}

The computational complexity divides into initialization costs incurred once and iteration costs incurred at each refinement step. The initialization constructs the $k$-nearest neighbor graph requiring $O(n^2 d)$ time for naive search or $O(n \log n)$ time using space-partitioning trees such as KD-trees or ball trees, where $d$ denotes ambient dimension. For high-dimensional data where $d > 20$, approximate nearest neighbor methods based on locality-sensitive hashing or random projection trees provide substantial speedup with negligible accuracy loss. The nerve complex construction requires checking $k$-cliques in the neighborhood graph, with worst-case complexity exponential in $k$ but typically linear in the number of edges for sparse graphs arising from proximity relationships.

The initial mass matrix computation requires $O(n k^2)$ time to evaluate all pairwise and triple intersection masses. When the complex includes triangles, computing the full edge mass matrix with off-diagonal entries requires $O(n_2)$ time where $n_2$ denotes the number of triangles. The Laplacian assembly through sparse matrix multiplication $L_0 = B_1^\top M_1^{-1} B_1$ requires $O(m^2)$ time where $m$ denotes the number of edges, though exploitation of sparsity typically reduces this to $O(m \bar{d})$ where $\bar{d}$ represents average vertex degree. The spectral decomposition using iterative methods such as Lanczos or Arnoldi requires $O(m \cdot p \cdot n_{\text{iter}})$ time where $p$ denotes the number of eigenvalues and $n_{\text{iter}}$ represents iterations to convergence, typically 50-200 iterations for moderate tolerance.

The iteration costs dominate overall computational burden. Each iteration performs density diffusion requiring $O(n p)$ time for matrix-vector products $V \exp(-t\Lambda) V^\top \rho_0^{(\ell)}$ with eigenvector matrix $V$. The mass matrix update requires $O(n k^2 + n_2)$ time to recompute intersection masses and modulate the full matrix. The response smoothing with GCV selection requires $O(J \cdot n p)$ time where $J$ denotes the number of grid points, typically $J \approx 20$-$30$. The Laplacian assembly requires $O(m^2)$ time for matrix operations. The total iteration cost scales as $O(n k^2 + m^2 + J \cdot n p)$ per iteration.

For typical datasets with $n \approx 1000$ observations, $k \approx 20$ neighbors, $p \approx 100$ spectral components, and 5-10 iterations, the dominant costs are the initial spectral decomposition and the repeated GCV-based response smoothing. The fixed covering strategy provides substantial computational advantages by maintaining constant topology, eliminating the need to rebuild the nerve complex or recompute the spectral decomposition across iterations. All geometric adaptation occurs through mass matrix evolution requiring only $O(n k^2)$ recomputation of intersection masses.

We address several potential sources of numerical instability through careful implementation. The edge mass matrix $M_1$ can become ill-conditioned when response modulation reduces some edge masses to near-zero values while others retain large masses. This creates large condition numbers requiring regularized inversion. We add a small ridge term $M_1 + \epsilon I$ with $\epsilon \approx 10^{-8} \cdot \|M_1\|_F$ when the condition number exceeds a threshold such as $10^8$, providing numerical stability while preserving geometric structure. Alternatively, we prune degenerate edges with mass below $10^{-6}$ of the maximum edge mass before inversion, effectively removing edges that have become infinitely long in the Riemannian metric.

The heat kernel $\exp(-t L_0)$ can produce numerical overflow or underflow when diffusion time $t$ or eigenvalues become extreme. We compute the exponential in the spectral domain as $\exp(-t\Lambda) = \text{diag}(\exp(-t\lambda_1), \ldots, \exp(-t\lambda_p))$, clamping very small exponentials to machine epsilon and very large exponentials to machine maximum. This prevents numerical artifacts while preserving essential spectral filtering structure. The clamping affects only extreme eigenvalues corresponding to very high-frequency modes that should be filtered aggressively anyway.

The GCV grid search can fail to find the optimal smoothing parameter when the criterion has multiple local minima or flat regions. We address this through logarithmic spacing that concentrates grid points where GCV varies most rapidly, typically in the range $[\epsilon, -\log(\epsilon)/\lambda_{\max}]$ where $\epsilon \approx 10^{-10}$. We also implement parabolic interpolation around the discrete minimum to refine the estimate, and we validate that the selected $\eta$ yields reasonable effective degrees of freedom (typically between 5 and $n/2$) to ensure the optimization succeeded.

When the Laplacian has multiple connected components, the smallest eigenvalue $\lambda_1 = 0$ corresponds to constant functions on each component. The heat kernel at this eigenvalue remains $\exp(0) = 1$, preserving component-wise constant functions. This creates no numerical difficulty but requires that we interpret smoothing as operating independently within each component. For regression, predictions remain constant within disconnected components if the response variation lies entirely in higher eigenspaces.

Spectral caching provides substantial efficiency gains for certain analysis workflows. After computing the initial eigendecomposition $L_0^{(0)} = V \Lambda V^\top$, we store the eigenvectors $V$ and eigenvalues $\Lambda$ for reuse. Bootstrap resampling for uncertainty quantification requires only re-smoothing the response with different bootstrap samples, reusing the same spectral decomposition and avoiding repeated expensive eigendecomposition. Similarly, multivariate response regression with $p$ response components requires $p$ smoothing operations but only one eigendecomposition, providing near-linear scaling in the number of responses.

The matrix exponential $\exp(-t L_0)$ factorizes as $V \exp(-t\Lambda) V^\top$, reducing the exponential to a diagonal matrix that can be computed element-wise. This enables efficient density evolution and response smoothing through matrix-vector products with $V$. The factorization also enables sensitivity analysis where we vary diffusion time $t$ without recomputing the full exponential, simply adjusting the diagonal entries $\exp(-t\lambda_i)$. For parameter tuning across multiple $t$ values, this provides orders-of-magnitude speedup.

\subsection*{4.4 Convergence Behavior and Diagnostics}

The iterative refinement exhibits predictable convergence patterns that provide
diagnostic information about geometric structure and parameter appropriateness.
We monitor three quantities at each iteration: relative change in smoothed
response
$$
\|\hat{y}^{(\ell)} - \hat{y}^{(\ell-1)}\| / \|\hat{y}^{(\ell-1)}\|,
$$
relative change in vertex density
$$
\|\rho_0^{(\ell)} - \rho_0^{(\ell-1)}\| / \|\rho_0^{(\ell-1)}\|,
$$
and relative change in edge density
$$
\|\rho_1^{(\ell)} - \rho_1^{(\ell-1)}\| / \|\rho_1^{(\ell-1)}\|.
$$
Convergence occurs when all three quantities fall below their respective
thresholds simultaneously, indicating geometric state stabilization.

Typical convergence occurs within 5-10 iterations for reasonable parameter
choices including $t \approx 1/\lambda_2$, $\beta \approx 0.1/t$, and
$\gamma \in [0.5, 2]$. Early iterations show large changes as geometry adapts
rapidly from the initial response-agnostic state toward response-aware
configurations. Middle iterations show decelerating changes as geometry
approaches equilibrium. Late iterations show negligible changes as fine-scale
adjustments stabilize. The response typically converges first within 3-5
iterations, followed by vertex densities within 5-8 iterations, with edge
densities sometimes requiring 8-12 iterations for full stabilization.

Slower convergence exceeding 15-20 iterations may indicate parameter mistuning
or highly irregular geometry requiring adaptive adjustment. Excessive damping
$\beta \gg 0.1/t$ prevents densities from evolving adequately, creating
oscillations as response modulation fights against overly constrained densities.
Insufficient damping $\beta \ll 0.1/t$ allows densities to concentrate
excessively in early iterations, creating instabilities where later modulation
cannot compensate. Extreme decay exponents $\gamma > 2$ or $\gamma < 0.5$
similarly create convergence difficulties through overly aggressive or overly
passive modulation.

Oscillating convergence where quantities decrease then increase alternately
suggests geometric-response feedback instability. This occurs when response
modulation creates sharp boundaries that density diffusion subsequently smooths,
causing geometry to oscillate between boundary-enhanced and boundary-depleted
states. Reducing diffusion time $t$ or increasing damping $\beta$ typically
stabilizes such oscillations by constraining geometric change magnitudes between
iterations. Alternatively, reducing decay exponent $\gamma$ makes modulation
less aggressive, allowing smoother adaptation.

Non-convergence after maximum iterations without obvious oscillation suggests
that the geometric state continues evolving without approaching equilibrium.
This may indicate that parameter choices create a dynamical system without
stable fixed points, requiring parameter adjustment to restore convergence. More
commonly, it reflects convergence thresholds remaining too strict for the
problem's inherent geometric complexity, where loosening thresholds by a factor
of two typically enables convergence without meaningful accuracy loss.

Diagnostic visualization provides insight into convergence behavior and
parameter appropriateness. Plotting vertex densities $\rho_0^{(\ell)}$ colored
on the feature space shows how probability mass concentrates in dense,
well-connected regions over iterations. Plotting edge masses $M_1^{(\ell)}[e,e]$
shows how connectivity patterns evolve with edges in response-coherent regions
retaining mass while boundary-crossing edges lose mass. Plotting smoothed
response $\hat{y}^{(\ell)}$ shows how conditional expectation estimates refine
from initial geometric smoothing to final response-aware prediction. These
visualizations reveal whether iteration produces meaningful geometric adaptation
or merely circulates without substantive improvement.

The convergence diagnostics also inform parameter selection. Rapid convergence
in under 3 iterations may indicate over-smoothing where parameters prevent
genuine geometric adaptation, suggesting reduced damping or increased decay
exponent. Very slow convergence exceeding 20 iterations with monotonic decrease
suggests under-smoothing where parameters create excessive geometric freedom,
suggesting increased damping or reduced decay exponent. The optimal parameter
regime typically produces convergence within 5-10 iterations with clear
geometric evolution visible in diagnostic plots but without oscillation or
instability.

When applying the framework to new datasets, we recommend the following
diagnostic workflow. Initialize with default parameters ($t = 1/\lambda_2$,
$\beta = 0.1/t$, $\gamma = 1$, uniform initial density) and run for 20
iterations monitoring convergence. If convergence occurs within 10 iterations
with stable behavior, the defaults are appropriate. If slower convergence or
oscillation occurs, adjust damping $\beta$ first as it most strongly affects
stability. If response boundaries appear too smooth or too sharp in
visualizations, adjust decay exponent $\gamma$ to control modulation strength.
This systematic approach typically identifies suitable parameters within 3-5
trial runs.

\section*{5. Discussion}

Geometric perspectives have long enriched scientific understanding. Karl Pearson
argued in The Grammar of Science that natural laws are fundamentally geometric
\cite{pearson1892grammar}, a view reflected in modern geometric data analysis
through spectral methods, manifold learning, and diffusion-based approaches.
Within this geometric framework, our contribution lies in the iterative refinement
of the analysis domain: we construct geometric objects that adapt to both feature
and response structure, allowing the domain of statistical estimation to evolve
rather than remain fixed.

\subsection*{5.1 Extensions and Generalizations}

The framework extends naturally to several important problem settings beyond
univariate regression. These extensions require minimal modification to the core
algorithm, demonstrating the flexibility of the geometric approach and its
applicability across diverse statistical tasks.

The multivariate response setting where $y \in \mathbb{R}^p$ represents a vector
of $p$ related outcomes measured on each observation requires no change to the
geometric state construction, which depends only on the feature matrix $X$. The
response smoothing adapts by applying the heat kernel to each response component
independently, producing $\hat{y}^{(\ell)} \in \mathbb{R}^{n \times p}$ where
each column represents a smoothed response component. The generalized
cross-validation criterion extends to multivariate form by summing squared
residuals across all components, yielding
$$
\text{GCV}(\eta) = \sum_{j=1}^p \|y_j - \hat{y}_j\|^2 / (n - \text{tr}(S_\eta))^2
$$
where the effective degrees of freedom apply uniformly to all components through
the shared geometric structure.

The response-coherence modulation for multivariate responses uses the Euclidean
norm of response differences. For an edge $[i,j]$ with endpoint response vectors
$$
\hat{y}^{(\ell)}(i), \hat{y}^{(\ell)}(j) \in \mathbb{R}^p,
$$
we compute
$$
\Delta_{ij} = \|\hat{y}^{(\ell)}(i) - \hat{y}^{(\ell)}(j)\|_2,
$$
measuring total response variation across all components. This provides a single
scalar measuring response coherence that generalizes naturally from the
univariate case. The penalty function $\Gamma(\Delta_{ij})$ proceeds as before
using the Euclidean norm, creating diffusion barriers where any response
component changes substantially. This approach treats all response components
symmetrically through the geometry, appropriate when components represent
related quantities measured in comparable units.

An alternative multivariate strategy considers response components
independently, computing separate penalty functions for each component and
combining them through geometric or arithmetic means. This allows different
response components to create barriers in different regions, potentially
improving prediction when components vary on different geometric scales or
exhibit distinct spatial structure. For instance, in microbiome applications
predicting both bacterial diversity and pH levels, the two responses may vary on
different length scales requiring component-specific modulation. However, this
increases parameter complexity and risks overfitting through excessive degrees
of freedom, suggesting that the symmetric approach provides a more robust
default.

The framework adapts to classification problems where $y$ takes discrete values
from a finite set. For binary classification with $y \in \{0,1\}$, we model the
conditional probability $P(y=1|X)$ as a smooth function on the geometric
complex. The smoothing proceeds as in regression, applying the heat kernel to
the binary response to produce $\hat{y}^{(\ell)} \in [0,1]$ representing
estimated class probabilities at each vertex. The response-coherence modulation
uses the absolute difference in estimated probabilities
$|\hat{y}^{(\ell)}(i) - \hat{y}^{(\ell)}(j)|$ to create barriers between regions
with different class distributions. Edges connecting vertices with similar
predicted probabilities retain high mass, creating diffusion corridors within
regions of homogeneous classification. Edges crossing class boundaries receive
low mass, creating barriers that sharpen the decision boundary.

For multiclass classification with $y \in \{1,\ldots,C\}$, we use one-hot
encoding to represent each observation's class as a vector in $\mathbb{R}^C$.
The framework treats this as a multivariate response with $C$ components,
applying heat kernel smoothing to each component independently. The
response-coherence modulation uses the Euclidean distance between one-hot
vectors, which equals $\sqrt{2}$ for observations in different classes and 0 for
observations in the same class. This creates strong barriers at class boundaries
while preserving smooth transitions within class regions where probabilities
vary continuously. The smoothed vectors
$\hat{y}^{(\ell)} \in \mathbb{R}^{n \times C}$ provide estimated class
probabilities after normalization, with prediction proceeding by selecting the
class with maximum probability at each vertex.

An alternative classification strategy uses the framework for geometric feature
extraction followed by standard classifiers. We run the iterative refinement to
convergence, extract the evolved vertex densities and diffusion distances as
features, then train a classifier such as logistic regression or support vector
machines on these geometric features rather than the original ambient
coordinates. This leverages the geometric adaptation while maintaining
compatibility with existing classification methodology and theoretical
guarantees. The evolved densities capture local sampling heterogeneity while
diffusion distances encode response-aware proximity, providing features that
respect the intrinsic structure discovered through iteration. This two-stage
approach separates geometric learning from classification, potentially improving
interpretability and enabling rigorous uncertainty quantification through
standard classification theory.

\subsection*{5.2 Connections to Related Work}

The iterative geometric framework synthesizes ideas from several areas of
statistical and computational methodology, establishing connections to manifold
learning, anisotropic diffusion, partition-based methods, and kernel approaches.
These connections clarify the method's intellectual lineage and position it
within the broader landscape of geometric data analysis.

The framework shares fundamental ideas with manifold learning and spectral
methods for dimensionality reduction. Methods such as Laplacian eigenmaps,
diffusion maps, and local tangent space alignment construct low-dimensional
representations by analyzing the spectral properties of graph Laplacians derived
from proximity relationships. Our approach differs by explicitly constructing a
Riemannian structure on the nerve complex rather than embedding data in
Euclidean space, and by iteratively refining the geometry to reflect response
structure rather than using fixed feature-based graphs. The connection to
diffusion maps proves particularly strong, as both methods use heat kernel
analysis to define geometry-respecting distances. However, diffusion maps
construct a single fixed distance from the initial geometry, while our framework
adapts the geometry itself through response-coherence modulation, producing
distances that evolve toward outcome-aware configurations.

The density evolution via heat diffusion with Tikhonov damping connects to
anisotropic diffusion and adaptive mesh refinement in numerical analysis and
image processing. Perona-Malik anisotropic diffusion smooths images while
preserving edges by making diffusion coefficients depend on local gradients,
creating barriers at boundaries. Our response-coherence modulation operates
analogously, adjusting edge masses based on response variation to create
diffusion barriers at outcome boundaries while preserving smoothing within
homogeneous regions. The mathematical structure differs in that Perona-Malik
diffusion modifies the diffusion equation itself, while we modulate the
Riemannian structure on which standard heat diffusion operates. This geometric
perspective provides theoretical clarity and computational advantages through
the spectral decomposition.

Adaptive mesh refinement in scientific computing iteratively adjusts
discretization grids to concentrate resolution where solutions vary rapidly,
analogous to our concentration of mass in response-coherent regions. The
connection extends to h-adaptivity in finite element methods where element sizes
adjust based on error indicators, and p-adaptivity where polynomial degrees vary
spatially. Our framework implements a form of geometric adaptivity where the
Riemannian structure concentrates or disperses mass, effectively adapting the
analysis domain to the regression function's smoothness. This perspective
suggests connections to a posteriori error estimation and adaptive
discretization theory that merit further investigation.

The relationship to partition-based methods including classification and
regression trees, random forests, and multivariate adaptive regression splines
illuminates complementary strengths and tradeoffs. These methods construct
disjoint partitions of feature space where the response varies little within
cells but changes substantially across cell boundaries. The partitions adapt
during training to reflect response structure, providing interpretable decision
rules and capturing sharp boundaries. Our framework achieves similar
response-awareness through geometry modulation but maintains overlapping
neighborhoods rather than disjoint cells, preserving smooth transitions and
capturing higher-dimensional simplicial structure. The partition view suggests
interpreting our method as constructing a "soft partition" where membership is
continuous and cells overlap substantially, generalizing the discrete partition
paradigm to a geometric setting.

Recent work on neural networks and deep learning reveals surprising connections
to geometric methods. Neural networks implicitly learn feature representations
that concentrate on lower-dimensional structures, with the learned
representations determining effective distances and neighborhoods. Our framework
makes this geometric learning explicit through the iterative refinement of the
nerve complex and its Riemannian structure. The connection suggests that neural
network success in regression tasks may partially derive from implicit geometric
adaptation, a hypothesis supported by recent theoretical work on the geometric
properties of learned representations. Our explicit geometric approach provides
interpretability and mathematical structure that neural methods lack, while
neural methods achieve greater flexibility through multiple layers and nonlinear
transformations.

Connections to Gaussian process regression and kernel methods clarify the
statistical interpretation of our spectral filtering approach. The heat kernel
$\exp(-\eta L_0)$ serves as a covariance kernel on the vertex set, with
correlation between vertices determined by heat flow patterns. This provides a
bridge to Gaussian process methodology where predictions combine kernel
smoothing with uncertainty quantification. However, standard Gaussian process
regression uses fixed kernels determined by distance functions in the ambient
space, while our framework adapts the kernel itself through geometric
refinement. The iterative evolution of the Laplacian produces a sequence of
kernels that progressively incorporate response information, yielding a form of
adaptive Gaussian process regression where the covariance structure evolves
toward outcome-aware configurations.

The framework also connects to semi-supervised learning and transductive
inference where labels are available for only a subset of observations. The
geometric structure constructed from all observations, both labeled and
unlabeled, provides a scaffold for propagating label information throughout the
feature space. The heat diffusion naturally implements label propagation, with
labels diffusing from labeled to unlabeled vertices according to the evolved
geometry. This connection suggests extensions where the iterative refinement
alternates between propagating labels and updating geometry, potentially
improving semi-supervised learning performance by discovering geometry that
respects both feature and label structure.

\subsection*{5.3 Theoretical Perspectives and Open Questions}

The iterative framework raises several fundamental theoretical questions about
convergence guarantees, consistency, and asymptotic properties that remain open
for future investigation. These questions connect to classical statistical
theory while introducing novel aspects arising from the geometric adaptation.

The convergence theory for the iterative scheme requires analysis of a coupled
dynamical system where densities evolve through heat diffusion while geometry
adapts through response-coherence modulation. Does the iteration always converge
to a unique fixed point or can multiple equilibria exist? The question parallels
classical fixed-point theorems but involves operators that are not contractive
in standard norms due to the nonlinear coupling between density and geometry.
Initial numerical experiments suggest convergence occurs reliably for reasonable
parameter choices, but establishing conditions guaranteeing convergence and
characterizing convergence rates as functions of the spectral gap and geometric
regularity remains an open problem.

The existence of multiple equilibria would have practical implications for
parameter selection and initialization strategy. If different initial
configurations lead to distinct stable configurations with different prediction
accuracy, the initialization becomes critical rather than merely convenient. The
empirical observation that uniform and distance-based initializations typically
converge to similar configurations suggests that the iteration may be relatively
insensitive to initialization, but this requires rigorous investigation.
Establishing uniqueness of the fixed point under appropriate conditions would
provide theoretical reassurance that the method produces reproducible results.

The statistical properties of the final predictor including consistency and
convergence rates require careful analysis connecting to nonparametric
regression theory. As sample size $n$ increases with the $k$-nearest neighbor
parameter $k$ growing appropriately, does the estimated conditional expectation
$\hat{y}^{(*)}$ converge to the true conditional expectation $E[y|X]$? The
question connects to extensive literature on $k$-nearest neighbor regression,
kernel smoothing, and spectral regularization, but the adaptive geometry
introduces novel aspects requiring new theoretical development. Standard
consistency proofs assume fixed kernels or neighborhoods determined by the
feature distribution alone, while our framework allows geometry to depend on
response observations.

The convergence rates in appropriate function spaces depend on both the
intrinsic dimension of the feature distribution and the smoothness of the
regression function with respect to the intrinsic geometry. For data
concentrating on a smooth $d$-dimensional manifold embedded in $\mathbb{R}^D$
with $d \ll D$, we expect convergence rates depending on $d$ rather than $D$,
providing the dimension reduction that motivates geometric methods. However, the
response-dependent geometry complicates the analysis because the relevant notion
of smoothness involves the evolved Riemannian structure rather than the ambient
or initial geometry. Establishing minimax optimal rates and comparing them to
classical methods represents an important open problem.

The geometric adaptation itself suggests connections to manifold learning theory
and geometric measure theory. The iterative refinement constructs a discrete
Riemannian manifold approximating the feature distribution with geometry adapted
to the response structure. As $n \to \infty$ with $k \to \infty$ appropriately,
does the discrete geometry converge to a limiting continuous Riemannian
manifold? This question connects to consistency results for graph Laplacians as
discretizations of Laplace-Beltrami operators, extended to the setting where the
metric itself undergoes adaptation. The limiting manifold, if it exists, would
depend on both the feature distribution and the response structure, representing
a geometric object encoding the statistical relationship between features and
outcomes.

The relationship between the limiting manifold and the data generating
distribution could be characterized through geometric measure theory. The
response-coherence modulation concentrates mass in regions where the response
varies smoothly, suggesting that the limiting metric assigns short distances
within response-homogeneous regions and long distances across response
boundaries. This creates a geometry where geodesics follow paths of constant
response value, analogous to level sets in gradient flow theory. Establishing
precise connections to geometric measure theory and characterizing the limiting
geometry as a functional of the joint distribution of features and response
would provide deep theoretical insight into the method's behavior.

The response-coherence modulation creates an adaptive geometry reminiscent of
adaptive metrics in variational problems and geometric flows. Theoretical
analysis could characterize limiting behavior as the decay exponent $\gamma$ and
diffusion time $t$ vary, potentially establishing phase transitions between
under-smoothed and over-smoothed regimes. The interplay between density
diffusion and response modulation suggests connections to reaction-diffusion
systems and pattern formation where similar feedback mechanisms create spatial
structures. The mathematical framework of geometric flows including Ricci flow
and mean curvature flow may provide analytical tools for studying the evolution
of our discrete Riemannian structures, establishing connections between
statistical methodology and differential geometry.

\subsection*{5.4 Practical Considerations and Future Directions}

The geometric framework proves most valuable for datasets exhibiting high
ambient dimension but low intrinsic dimension, where measurements involve many
variables but genuine variation concentrates on lower-dimensional structures.
Microbiome compositions exemplify this setting, with hundreds or thousands of
bacterial taxa yielding compositional vectors that cluster along manifolds
determined by ecological constraints. Genomic expression profiles, chemical
reaction networks, and spatial point patterns exhibit similar structure, making
them natural application domains. The method requires sufficient sample size to
populate the intrinsic structure adequately, typically $n > 100$ for two or
three-dimensional intrinsic structures, with sample size requirements growing
exponentially with intrinsic dimension as expected from manifold sampling
theory.

When the data do not exhibit intrinsic low-dimensional structure or the response
varies independently of any underlying geometry, the framework provides little
advantage over classical regularization methods. High-dimensional data uniformly
distributed in the ambient space without clustering or concentration lack the
geometric structure that the method exploits. Similarly, if the response depends
on many ambient coordinates without respecting intrinsic geometry, attempting to
adapt geometry to response structure may produce overfitting rather than useful
regularization. Preliminary diagnostics examining nearest neighbor distance
distributions and spectral properties of initial graph Laplacians can assess
whether geometric methods are appropriate before committing computational
resources to the full iterative scheme.

Computational limitations constrain the method's applicability to datasets with
more than approximately 10,000 observations using standard implementations. The
spectral decomposition requiring eigenvalue computation for sparse Laplacian
matrices becomes expensive as problem size grows, with costs scaling roughly as
$O(m \cdot p)$ where $m$ denotes edge count and $p$ denotes the number of
eigenvalues. Larger datasets require approximate methods including randomized
spectral decomposition, multilevel preconditioning, or divide-and-conquer
strategies that partition the data into smaller geometric units processed
separately. Recent advances in scalable spectral methods and the development of
specialized algorithms for geometric graphs suggest that extensions to much
larger datasets may become feasible with continued algorithmic development.

The framework provides uncertainty quantification through Bayesian credible
intervals computed in the spectral domain. The posterior distribution for the
regression function follows from the Tikhonov regularization formulation, where
the smoothed response represents the posterior mean and the regularization
parameter determines the posterior precision. Given the spectral decomposition
$L_0 = V\Lambda V^\top$ and residual variance estimate $\hat{\sigma}^2$, the
posterior standard deviation for each spectral coefficient follows from the
regularization formula
$$
\text{SD}(\alpha_j | y) = \hat{\sigma} / \sqrt{1 + \eta \lambda_j}.
$$
Drawing Monte Carlo samples from these independent normal posteriors in the
spectral domain and transforming to the vertex domain via
$$
\hat{y}_{\text{sample}} = V \alpha_{\text{sample}}
$$
produces samples from the posterior distribution of the regression function.
Empirical quantiles of these samples provide pointwise credible intervals at each
vertex with specified coverage probability. This approach leverages the spectral
structure to provide computationally efficient uncertainty quantification without
requiring bootstrap resampling or iterative MCMC methods.

The methods described in this paper are implemented in the R package
\texttt{gflow} (Gradient Flow for Regression), available at
\url{https://github.com/pgajer/gflow}. The package provides both core
algorithmic modules and visualization tools for exploratory analysis, using C++
for computational efficiency while maintaining an intuitive R interface for data
analysis workflows. The implementation includes the initialization procedures,
iterative refinement cycle, parameter selection strategies, and diagnostic tools
described in Section 4.

Future development directions include extending the framework to longitudinal
data where observations consist of time series or repeated measurements,
incorporating temporal structure into the geometric refinement. The nerve
complex could record temporal connections in addition to spatial proximity, with
density diffusion operating on the combined space-time structure. Another
direction involves hierarchical extensions where the geometric state organizes
features at multiple resolutions, producing nested complexes that capture
structure at different scales simultaneously. This connects to multiscale
geometric analysis and wavelet methods, potentially improving performance when
the response exhibits structure at multiple scales.

The theoretical questions outlined in Section 5.3 represent a substantial
research program connecting statistical methodology, differential geometry, and
numerical analysis. Establishing convergence guarantees, consistency results,
and optimal rates would place the framework on firm theoretical foundations.
Characterizing the limiting geometry and its relationship to the data generating
distribution through geometric measure theory would provide deep insight into
how geometric adaptation operates. These theoretical advances would complement
the empirical success of the methods and guide further methodological
development.

The geometric perspective on statistical analysis embodied in this framework
reflects a broader movement toward explicitly geometric and
differential-geometric approaches in data science. As datasets grow in
complexity and structure, methods that adapt to intrinsic geometry rather than
imposing ambient coordinate systems provide increasingly valuable tools. The
iterative refinement scheme demonstrates how classical ideas from differential
geometry, spectral theory, and diffusion processes can synthesize into practical
methodology for statistical analysis, establishing connections between pure
mathematics and applied data science that benefit both domains.

\bibliographystyle{plain}
\bibliography{no12_bibliography}

\begin{thebibliography}{10}

\bibitem{barbarossa2020topological}
Sergio Barbarossa and Stefania Sardellitti.
\newblock Topological signal processing over simplicial complexes.
\newblock {\em IEEE Transactions on Signal Processing}, 68:2992--3007, 2020.

\bibitem{belkin2003laplacian}
Mikhail Belkin and Partha Niyogi.
\newblock Laplacian eigenmaps for dimensionality reduction and data
  representation.
\newblock {\em Neural Computation}, 15(6):1373--1396, 2003.

\bibitem{belkin2006manifold}
Mikhail Belkin, Partha Niyogi, and Vikas Sindhwani.
\newblock Manifold regularization: A geometric framework for learning from
  labeled and unlabeled examples.
\newblock {\em Journal of Machine Learning Research}, 7:2399--2434, 2006.

\bibitem{borsuk1948imbedding}
Karol Borsuk.
\newblock On the imbedding of systems of compacta in simplicial complexes.
\newblock {\em Fundamenta Mathematicae}, 35(1):217--234, 1948.

\bibitem{breiman1984classification}
Leo Breiman, Jerome~H. Friedman, Richard~A. Olshen, and Charles~J. Stone.
\newblock {\em Classification and Regression Trees}.
\newblock Wadsworth International Group, Belmont, CA, 1984.

\bibitem{callahan2017replication}
Benjamin~J Callahan, Daniel~B DiGiulio, Daniela~SA Goltsman, Cherie~L Sun,
  Elizabeth~K Costello, Pratheepa Jeganathan, Joseph~R Biggio, Ronald~J Wong,
  Maurice~L Druzin, Gary~M Shaw, et~al.
\newblock Replication and refinement of a vaginal microbial signature of
  preterm birth in two racially distinct cohorts of us women.
\newblock {\em Proceedings of the National Academy of Sciences},
  114(37):9966--9971, 2017.

\bibitem{callahan2016dada2}
Benjamin~J Callahan, Paul~J McMurdie, Michael~J Rosen, Andrew~W Han, Amy Jo~A
  Johnson, and Susan~P Holmes.
\newblock {DADA2}: High-resolution sample inference from illumina amplicon
  data.
\newblock {\em Nature Methods}, 13(7):581--583, 2016.

\bibitem{carlsson2009topology}
Gunnar Carlsson.
\newblock Topology and data.
\newblock {\em Bulletin of the American Mathematical Society}, 46(2):255--308,
  2009.

\bibitem{chaudhuri2014rates}
Kamalika Chaudhuri and Sanjoy Dasgupta.
\newblock Rates of convergence for nearest neighbor classification.
\newblock {\em The Annals of Statistics}, 42(1):1--30, 2014.

\bibitem{chung1997spectral}
Fan~RK Chung and F~Chung Graham.
\newblock {\em Spectral graph theory}.
\newblock Number~92 in CBMS Regional Conference Series in Mathematics. American
  Mathematical Society, 1997.

\bibitem{coifman2006diffusion}
Ronald~R Coifman and St{\'e}phane Lafon.
\newblock Diffusion maps.
\newblock {\em Applied and computational harmonic analysis}, 21(1):5--30, 2006.

\bibitem{desbrun2005discrete}
Mathieu Desbrun, Anil~N Hirani, Melvin Leok, and Jerrold~E Marsden.
\newblock Discrete exterior calculus.
\newblock {\em arXiv preprint math/0508341}, 2005.

\bibitem{dowker1952homology}
Clifford~Hugh Dowker.
\newblock Homology groups of relations.
\newblock {\em Annals of Mathematics}, pages 84--95, 1952.

\bibitem{edelsbrunner2010computational}
Herbert Edelsbrunner and John Harer.
\newblock {\em Computational Topology: An Introduction}.
\newblock American Mathematical Society, Providence, RI, 2010.

\bibitem{elovitz2019cervicovaginal}
Michal~A Elovitz, Pawel Gajer, Virginia Riis, Amy~G Brown, Michael~S Humphrys,
  Johanna~B Holm, and Jacques Ravel.
\newblock Cervicovaginal microbiota and local immune response modulate the risk
  of spontaneous preterm delivery.
\newblock {\em Nature Communications}, 10(1):1305, 2019.

\bibitem{fefferman2016testing}
Charles Fefferman, Sanjoy Mitter, and Hariharan Narayanan.
\newblock Testing the manifold hypothesis.
\newblock {\em Journal of the American Mathematical Society}, 29:983--1049,
  2016.

\bibitem{fettweis2019vaginal}
Jennifer~M Fettweis, Myrna~G Serrano, J~Paul Brooks, David~J Edwards,
  Philippe~H Girerd, Hardik~I Parikh, Bernice Huang, Tomasz~J Arodz, Lavanya
  Edupuganti, Abigail~L Glascock, et~al.
\newblock The vaginal microbiome and preterm birth.
\newblock {\em Nature Medicine}, 25(6):1012--1021, 2019.

\bibitem{friedman1991multivariate}
Jerome~H. Friedman.
\newblock Multivariate adaptive regression splines.
\newblock {\em The Annals of Statistics}, 19(1):1--67, 1991.

\bibitem{gajer2025geometry}
Pawel Gajer and Jacques Ravel.
\newblock The geometry of machine learning models.
\newblock {\em arXiv preprint arXiv:2508.02080}, 2025.

\bibitem{gerber2012morse}
Samuel Gerber and Ross Whitaker.
\newblock Morse-smale regression.
\newblock {\em Journal of Computational and Graphical Statistics},
  22(1):193--214, 2013.

\bibitem{ghrist2008barcodes}
Robert Ghrist.
\newblock Barcodes: the persistent topology of data.
\newblock {\em Bulletin of the American Mathematical Society}, 45(1):61--75,
  2008.

\bibitem{giusti2016two}
Chad Giusti, Robert Ghrist, and Danielle~S Bassett.
\newblock Two's company, three (or more) is a simplex.
\newblock {\em Journal of Computational Neuroscience}, 41(1):1--14, 2016.

\bibitem{goodfellow2016deep}
Ian Goodfellow, Yoshua Bengio, and Aaron Courville.
\newblock {\em Deep learning}.
\newblock MIT press, 2016.

\bibitem{grady2010discrete}
Leo~J Grady and Jonathan~R Polimeni.
\newblock Discrete calculus applied to modeling of tumor growth.
\newblock {\em Discrete Calculus: Applied Analysis on Graphs for Computational
  Science}, pages 273--280, 2010.

\bibitem{grigoryan2009heat}
Alexander Grigor'yan.
\newblock {\em Heat Kernel and Analysis on Manifolds}.
\newblock American Mathematical Society, Providence, RI, 2009.

\bibitem{hastie2015statistical}
Trevor Hastie, Robert Tibshirani, and Martin Wainwright.
\newblock {\em Statistical Learning with Sparsity: The Lasso and
  Generalizations}.
\newblock CRC Press, Boca Raton, FL, 2015.

\bibitem{hein2007graph}
Matthias Hein, Jean-Yves Audibert, and Ulrike Von~Luxburg.
\newblock From graphs to manifolds--weak and strong pointwise consistency of
  graph laplacians.
\newblock {\em Learning theory}, pages 470--485, 2007.

\bibitem{hirani2003discrete}
Anil~N Hirani.
\newblock {\em Discrete exterior calculus}.
\newblock PhD thesis, California Institute of Technology, 2003.

\bibitem{jones2008manifold}
Peter~W Jones, Mauro Maggioni, and Raanan Schul.
\newblock Manifold parametrizations by eigenfunctions of the laplacian and heat
  kernels.
\newblock {\em Proceedings of the National Academy of Sciences},
  105(6):1803--1808, 2008.

\bibitem{lafon2006diffusion}
St{\'e}phane Lafon and Ann~B Lee.
\newblock Diffusion maps and coarse-graining: A unified framework for
  dimensionality reduction, graph partitioning, and data set parameterization.
\newblock {\em IEEE Transactions on Pattern Analysis and Machine Intelligence},
  28(9):1393--1403, 2006.

\bibitem{levina2004maximum}
Elizaveta Levina and Peter~J Bickel.
\newblock Maximum likelihood estimation of intrinsic dimension.
\newblock {\em Advances in Neural Information Processing Systems}, 17, 2004.

\bibitem{li2007nonparametric}
Qi~Li and Jeffrey~S Racine.
\newblock {\em Nonparametric Econometrics: Theory and Practice}.
\newblock Princeton University Press, Princeton, NJ, 2007.

\bibitem{lim2020hodge}
Lek-Heng Lim.
\newblock Hodge laplacians on graphs.
\newblock {\em SIAM Review}, 62(3):685--715, 2020.

\bibitem{lovasz1993random}
L{\'a}szl{\'o} Lov{\'a}sz.
\newblock Random walks on graphs: A survey.
\newblock In {\em Combinatorics, Paul Erd{\H{o}}s is Eighty}, volume~2, pages
  1--46. J{\'a}nos Bolyai Mathematical Society, Budapest, 1993.

\bibitem{nadler2006diffusion}
Boaz Nadler, St{\'e}phane Lafon, Ronald~R Coifman, and Ioannis~G Kevrekidis.
\newblock Diffusion maps, spectral clustering and reaction coordinates of
  dynamical systems.
\newblock {\em Applied and Computational Harmonic Analysis}, 21(1):113--127,
  2006.

\bibitem{pearson1892grammar}
Karl Pearson.
\newblock {\em The Grammar of Science}.
\newblock Walter Scott, London, 1892.

\bibitem{petri2014homological}
Giovanni Petri, Paul Expert, Federico Turkheimer, Robin Carhart-Harris, David
  Nutt, Peter~J Hellyer, and Francesco Vaccarino.
\newblock Homological scaffolds of brain functional networks.
\newblock {\em Journal of The Royal Society Interface}, 11(101):20140873, 2014.

\bibitem{richards2013geometry}
Daniel Richards and Daniel~H Huson.
\newblock Geometry of the space of phylogenetic trees.
\newblock {\em Systematic Biology}, 62(4):549--564, 2013.

\bibitem{rinaldo2010stability}
Alessandro Rinaldo and Larry Wasserman.
\newblock Stability of density-based clustering.
\newblock {\em The Annals of Statistics}, 38(5):2998--3023, 2010.

\bibitem{schaub2021random}
Michael~T Schaub, Austin~R Benson, Paul Horn, Gabor Lippner, and Ali Jadbabaie.
\newblock Random walks on simplicial complexes and the normalized hodge
  1-laplacian.
\newblock {\em SIAM Review}, 62(2):353--391, 2020.

\bibitem{scott2015multivariate}
David~W Scott.
\newblock {\em Multivariate Density Estimation: Theory, Practice, and
  Visualization}.
\newblock John Wiley \& Sons, Hoboken, NJ, 2nd edition, 2015.

\bibitem{silverman1986density}
Bernard~W Silverman.
\newblock {\em Density Estimation for Statistics and Data Analysis}.
\newblock Chapman and Hall, London, 1986.

\bibitem{singer2006graph}
Amit Singer.
\newblock Graph diffusion distance and local clustering algorithms.
\newblock {\em Physical Review Letters}, 97(4):048701, 2006.

\bibitem{singer2009non}
Amit Singer and Ronald~R Coifman.
\newblock Non-linear independent component analysis with diffusion maps.
\newblock {\em Applied and Computational Harmonic Analysis}, 25(2):226--239,
  2008.

\bibitem{sizemore2018knowledge}
Ann~E Sizemore, Elisabeth~A Karuza, Chad Giusti, and Danielle~S Bassett.
\newblock Knowledge gaps in the early growth of semantic feature networks.
\newblock {\em Nature Human Behaviour}, 2(9):682--692, 2018.

\bibitem{stone1982optimal}
Charles~J Stone.
\newblock Optimal global rates of convergence for nonparametric regression.
\newblock {\em The Annals of Statistics}, 10(4):1040--1053, 1982.

\bibitem{szlam2008diffusion}
Arthur~D Szlam, Mauro Maggioni, and Ronald~R Coifman.
\newblock Diffusion-driven multiscale analysis on manifolds and graphs:
  top-down and bottom-up constructions.
\newblock {\em Wavelets XI}, 5914:445--455, 2005.

\bibitem{tibshirani1996regression}
Robert Tibshirani.
\newblock Regression shrinkage and selection via the lasso.
\newblock {\em Journal of the Royal Statistical Society Series B: Statistical
  Methodology}, 58(1):267--288, 1996.

\bibitem{ting2011analysis}
Daniel Ting, Ling Huang, and Michael~I Jordan.
\newblock An analysis of the convergence of graph laplacians.
\newblock {\em arXiv preprint arXiv:1101.5435}, 2011.

\bibitem{trillos2016variational}
Nicol{\'a}s~Garc{\'i}a Trillos and Dejan Slepčev.
\newblock Variational problems and pdes on implicit surfaces.
\newblock {\em Journal de Math{\'e}matiques Pures et Appliqu{\'e}es},
  105(6):743--788, 2016.

\bibitem{tsybakov2009introduction}
Alexandre~B Tsybakov.
\newblock {\em Introduction to Nonparametric Estimation}.
\newblock Springer, 2009.

\bibitem{vonluxburg2007tutorial}
Ulrike Von~Luxburg.
\newblock A tutorial on spectral clustering.
\newblock {\em Statistics and Computing}, 17(4):395--416, 2007.

\bibitem{vonluxburg2008consistency}
Ulrike Von~Luxburg, Mikhail Belkin, and Olivier Bousquet.
\newblock Consistency of spectral clustering.
\newblock {\em The Annals of Statistics}, 36(2):555--586, 2008.

\bibitem{wand1994kernel}
Matt~P Wand and M~Chris Jones.
\newblock {\em Kernel Smoothing}.
\newblock Chapman and Hall/CRC, 1994.

\bibitem{wasserman2006all}
Larry Wasserman.
\newblock {\em All of Nonparametric Statistics}.
\newblock Springer, New York, 2006.

\bibitem{zhou2004learning}
Dengyong Zhou, Olivier Bousquet, Thomas~N Lal, Jason Weston, and Bernhard
  Sch{\"o}lkopf.
\newblock Learning with local and global consistency.
\newblock In {\em Advances in Neural Information Processing Systems}, pages
  321--328, 2004.

\bibitem{zhu2003semi}
Xiaojin Zhu, Zoubin Ghahramani, and John~D Lafferty.
\newblock Semi-supervised learning using gaussian fields and harmonic
  functions.
\newblock In {\em Proceedings of the 20th International Conference on Machine
  Learning (ICML-03)}, pages 912--919. AAAI Press, 2003.

\end{thebibliography}

\end{document}